\RequirePackage{fix-cm}
\documentclass[smallextended]{svjour3}       
\smartqed  
\journalname{}

\usepackage{graphicx}
\usepackage[numbers,sort&compress]{natbib}
\usepackage{amsmath,amssymb,amsfonts}
\RequirePackage{doi}
\usepackage{hyperref}
\hypersetup{
	colorlinks=true,
	linkcolor=blue,
	filecolor=magenta,      
	urlcolor=cyan,
}
\usepackage[utf8]{inputenc}
\usepackage[noend]{algpseudocode}
\usepackage{algorithm}
\usepackage{enumitem}
\usepackage{listings}
\usepackage{rotating}
\usepackage{afterpage}
\usepackage{xcolor,colortbl}

\begin{document}

\title{A Model-Driven Approach to Machine Learning and Software Modeling for the IoT 
}
\subtitle{Generating Full Source Code for Smart Internet of Things (IoT) Services and Cyber-Physical Systems (CPS)}

\titlerunning{MDE for ML and Software}        

\author{Armin Moin \and
        Moharram Challenger \and
		Atta Badii \and
        Stephan G{\"u}nnemann
}

\authorrunning{Moin et al.} 

\institute{Armin Moin (corresponding author) \at
            Department of Informatics, Technical University of Munich, Germany \\
            \email{moin@in.tum.de}           
            \and
            Moharram Challenger \at
            Department of Computer Science, University of Antwerp \& Flanders Make, Belgium \\
            \email{moharram.challenger@uantwerpen.be} 
            \and
            Atta Badii \at
            Department of Computer Science, University of Reading, United Kingdom \\
            \email{atta.badii@reading.ac.uk}           
            \and
            Stephan G{\"u}nnemann \at
            Department of Informatics \& Munich Data Science Institute, Technical University of Munich, Germany \\
            \email{guennemann@in.tum.de}           
}

\date{Received: date / Accepted: date}

\maketitle

\begin{abstract}
Models are used in both Software Engineering (SE) and Artificial Intelligence (AI). SE models may specify the architecture at different levels of abstraction and for addressing different concerns at various stages of the software development life-cycle, from early conceptualization and design, to verification, implementation, testing and evolution. However, AI models may provide smart capabilities, such as prediction and decision-making support. For instance, in Machine Learning (ML), which is currently the most popular sub-discipline of AI, mathematical models may learn useful patterns in the observed data and can become capable of making predictions. The goal of this work is to create synergy by bringing models in the said communities together and proposing a holistic approach to model-driven software development for intelligent systems that require ML. We illustrate how software models can become capable of creating and dealing with ML models in a seamless manner. The main focus is on the domain of the Internet of Things (IoT), where both ML and model-driven SE play a key role. In the context of the need to take a Cyber-Physical System-of-Systems perspective of the targeted architecture, an integrated design environment for both SE and ML sub-systems would best support the optimization and overall efficiency of the implementation of the resulting system. In particular, we implement the proposed approach, called ML-Quadrat, based on ThingML, and validate it using a case study from the IoT domain, as well as through an empirical user evaluation. It transpires that the proposed approach is not only feasible, but may also contribute to the performance leap of software development for smart Cyber-Physical Systems (CPS) which are connected to the IoT, as well as an enhanced user experience of the practitioners who use the proposed modeling solution.

\end{abstract}

\section{Introduction}\label{introduction}
As software and information/data-intensive systems, such as Cyber-Physical Systems (CPS), which are highly complex systems of systems \cite{GeisbergerBroy2014}, become smarter through incorporating Artificial Intelligence (AI), and more pervasive via the Internet of Things (IoT) with billions of networked devices \cite{Atzori+2010}, we  observe an increasing need for integration and liaison between the Software and Systems Engineering (SSE) community on the one side and the AI, including the Data Analytics and Machine Learning (DAML) community on the other side. To this aim, two research directions motivated by the following broad research questions are evolving simultaneously: (i) How to enhance SSE through AI (e.g., DAML)? For instance, the field of Mining Software Repositories (MSR), which deals with applying DAML methods and techniques to large amounts of data that are stored in various formats in the software source code and bug repositories, in order to make software development more efficient, serves as an example for this direction. (ii) How can AI, e.g., DAML benefit from SSE approaches and paradigms, such as Model-Driven Software Engineering (MDSE), also known as Model-Based Software Engineering (MBSE)? This work lies at the intersection of the said research directions since it aims to bring both communities together and contribute to each one.

Due to the abstraction and the automation that they can provide, software models in the context of the MDSE paradigm, especially the Domain-Specific Modeling (DSM) methodology \cite{KellyTolvanen2008} with full code generation play an important role in the highly complex and very large software systems of today. In particular, in the IoT domain, where distributed systems with heterogeneous hardware and software platforms, programming languages and communication protocols are the norm, one can better perceive the additional value of models and domain-specific MDSE \cite{Schaetz2014, Harrand+2016}. Prior work in the literature, such as ThingML \cite{Morin+2017, Harrand+2016, Fleurey+2011, ThingML}, HEADS \cite{Morin+2016, HEADS} and $\mu$-Kevoree \cite{Fouquet+2012} (see Section \ref{related-work}) concentrated on domain-specific MDSE for the IoT/CPS domain. However, the main shortcoming of these models is that they cannot support the ever-increasing DAML requirements of software systems, in particular, in the IoT/CPS domain where massive datasets and data streams are being generated by the sensors and other devices. We argue that the Domain-Specific Modeling Languages (DSML) for the IoT/CPS have to include DAML concepts and offer access to the APIs of libraries and frameworks for DAML on the modeling layer. Otherwise, the DAML functionalities of smart, data-driven IoT Services and CPS applications need to be implemented separately either in a manual way or using other \lq{}silo\rq{} DSMLs. However, this would be in contrast to the DSM goal concerning being able to generate every artifact out of the abstract MDSE models, through model-to-code/model-to-text and model-to-model transformations, in an automated, integrated and seamless manner.

Furthermore, DAML models, such as Probabilistic Graphical Models (PGM) or Artificial Neural Networks (ANN) are not capable of acting as software models for the entire system, e.g., for modeling a complete smart IoT service or smart CPS application. Bishop \cite{Bishop2013} proposed Infer.NET \cite{InferNet}, which was a DSML, based on the probabilistic programming paradigm, in order to treat PGMs as both ML models and software models in the sense of domain-specific MDSE, where the entire software solution is generated out of the model. However, the main drawback of such an approach is that PGMs and other families of ML models are not expressive enough to be capable of modeling the entire system for IoT and CPS use case scenarios. 

In contrast, we enhance software models, in order to make them capable of creating, training, deploying and re-training ML models as necessary for IoT use cases. However, the proposed approach is not tied to any specific vertical problem (application) domain. This means, the proposed solution can be deployed in diverse vertical domains, such as smart healthcare and smart energy systems. This is in accordance with the nature of CPS, which are cross-domain by definition, and the IoT, that is an interconnection of all such cross-domain systems of systems \cite{Schaetz2014, GeisbergerBroy2014}. 

The original idea was proposed previously in our position paper \cite{Moin+2018}, as well as our poster/extended abstract \cite{Moin+2020}. In this work, we elaborate on the proposed approach more thoroughly, illustrate our implementation of the prototype that serves as the proof-of-concept, as well as the validation of the proposed approach. Hence, the contribution of this paper is twofold: (i) We validate the research hypothesis that software developers using the MDSE paradigm, particularly the DSM methodology, may have their software models enhanced with the capability to automatically produce and train ML models, and deal with them. Simultaneously, we maintain the feasibility of full source code generation in an automated way. The said ML models may affect the behavioral models of software systems. This is validated using a case study. (ii) In addition to the feasibility of the proposed approach, we validate the hypothesis that it contributes to the performance leap of software development in the IoT domain and leads to a higher level of satisfaction regarding the user experience of the practitioners (i.e., software developers, data scientists, etc.) who use the proposed approach. This is validated through an empirical evaluation by a number of external experts.

We provide our open-source prototype, called ML-Quadrat, with sufficient documentation and samples to facilitate using this as a platform to let both software developers and ML practitioners support new IoT platforms and ML libraries. This shall lead to open innovations and generate synergies in both the SSE and AI communities. Using the proposed approach, the SSE community is empowered with the state-of-the-art ML methods and techniques out-of-the-box, while the ML community can obtain access to the scalable, robust and efficient Software Engineering (SE) solutions, based on best practice. The integration of the said models from SE and ML is conducted in a seamless manner that does not require any knowledge and skills in the particular APIs of the underlying platforms and libraries. For instance, to generate Python code for ML, based on the APIs of different libraries and frameworks, one does not need to be familiar with their specific APIs. Our DSML abstracts from those platform-specific APIs, hence offering a higher layer of abstraction, i.e., the modeling layer. Different model-to-code transformations, also known as code generators can generate the entire source code for various DAML libraries and frameworks, e.g., Scikit-Learn \cite{Pedregosa+2011} and Keras \cite{Chollet+2015} with the TensorFlow \cite{Abadi+2015} backend in a fully automated manner. 

Moreover, since our work is built based on the open-source ThingML \cite{ThingML} project, we also inherit their code generators (\lq{}compilers\rq{}) for various platforms, programming languages and protocols. ThingML \cite{ThingML} can generate code in Java, C (Posix, Teensy, Arduino), C++, Javascript and Go. Further, they support not only the Hypertext Transfer Protocol (HTTP), but also the more suitable application layer communication protocols for resource-constrained IoT-devices, namely the Constrained Application Protocol (CoAP) for one-to-one communications and the Message Queuing Telemetry Transport (MQTT) protocol for many-to-many communications following the publish-subscribe pattern. In this work, we extend their approach, including the meta-model, as well as the code generation framework to enable generating Python code for supporting the required DAML functionalities.

The rest of this paper is structured as follows: Section \ref{background} provides the required background on the IoT/CPS and the preliminaries on analytics modeling, as well as software modeling. Moreover, Section \ref{related-work} reviews the state of the art and points out the gap in the literature that is being addressed by the present work. We propose our novel approach in Section \ref{proposed-approach} that is followed by presenting the open-source prototype in Section \ref{open-source-prototype}. Further, we validate the above-mentioned research hypotheses in Section \ref{validation}. Finally, we conclude and suggest future work in Section \ref{conclusion-futurework}.

\section{Background}\label{background}

\subsection{The Internet of Things (IoT) and Cyber-Physical Systems (CPS)}\label{background-iot-cps}

The original \textit{World Wide Web (WWW)} was developed in 1989 to enable automated information-sharing between scientists in universities and institutes around the globe \cite{CERN}. The term \textit{Web 2.0} was introduced in 1999 \cite{DiNucci1999} as user-generated content on the web gained more attention. In 2001, \textit{Web 3.0} or the \textit{Semantic Web} was introduced \cite{Berners-LeeHendler2001}. This was an extension of the web to support machine-readable multi-media content development on the web, i.e., \textit{semantic} data that could be processed and \textit{understood} by computers, such that they can conduct reasoning supported by the semantic markup. To this aim, the World Wide Web Consortium (W3C) promoted a set of standards, such as the Resource Description Framework (RDF) that could enable data from heterogeneous sources to be shared and reused across applications, websites and mobile apps. These semantic technologies let concepts, objects and their relationships be formally represented through meta-data, e.g., via ontologies. Today, the fourth generation of the web, i.e., \textit{Web 4.0}, which is better known as the \textit{Internet of Things (IoT)}, is being gradually formed. The IoT is an expansion of the Internet into new domains, devices and objects (i.e., \textit{things}), such as Radio-Frequency Identification (RFID) tags, sensors, actuators, mobile phones, etc. which through unique addressing schemes are able to interact and perhaps also cooperate with each other to reach common goals \cite{Atzori+2010}.

Another related but slightly different notion is Cyber-Physical Systems (CPS). Similar to the IoT systems, CPS, which are highly complex systems of systems that possess both physical and virtual (cyber) components \cite{GeisbergerBroy2014}, consist of heterogeneous and distributed platforms, such as various embedded micro-controllers. As more CPS are being connected to the Internet (IoT), we no longer need to distinguish between the two notions of CPS and IoT. Nevertheless, there is no consensus on the exact definition of CPS and its borders and/or possible overlaps with the IoT. The US National Institute of Standards and Technology (NIST) Special Publication on CPS and the IoT \cite{Greer+2019} also highlighted this fact and pointed out that CPS and the IoT have \lq{}distinct origins but overlapping definitions, with both referring to trends in integrating digital capabilities, including network connectivity and computational capability, with physical devices and systems\rq{}.

CPS have by nature special capabilities, known as the so-called \textit{cross-*}, \textit{live-*} and \textit{self-*} capabilities. The \textit{cross-*} capabilities, include cross-domain, cross-technology, cross-organization and cross-functional. Moreover, the \textit{live-*} capabilities comprise live-re-configuration, live-re-deployment, live-update, live-enhancement and live-extension. Further, the \textit{self-*} capabilities are self- documenting, self-monitoring/diagnosis, self-optimizing, self- healing and self-adapting/training \cite{Schaetz2014}. 

Since CPS involve both the physical and the virtual (cyber/digital) worlds, modeling them is quite challenging. For instance, in the physical world, the dynamics of the system is captured by a set of variables that change their values \emph{continuously} over time. The dependencies between these variables are captured by \emph{continuous functions} that are expressed by differential calculus and integration theory, where time is represented by real numbers. By contrast, digital systems can be modeled as discrete event systems with a number of states. They can be modeled, e.g., via state machines or Petri-Nets. Thus, in digital systems, time is discrete. Furthermore, in such systems, the notion of causality, i.e., the logical dependencies between the events, might be more sophisticated than the notion of time \cite{Broy2008, GeisbergerBroy2014}. Finally, Papatheocharous et al. \cite{Papatheocharous+2013} proposed a closely related and similar concept to CPS in their position paper, called Federated Embedded Systems (FES).

In this work, we focus on modeling IoT services that require smart capabilities through Machine Learning (ML). As a motivating example, let us consider a condition-based monitoring system of a hydraulics system in an industrial facility. The goal is to conduct predictive maintenance via ML models that are trained on the data that are acquired from a number of various sensors (e.g., multiple pressure and temperature sensors), thus enabling the prediction of possible future faults of the system by the ML models. Moreover, there exist a number of \textit{virtual sensors}, whose values are not directly measured by any physical sensor device, but they are calculated based on other sensor measurements. One example is the cooling efficiency. There is no sensor to measure this quantity explicitly, but it is calculated according to the oil temperature at the cooler, one of the temperature sensors, as well as the ambient temperature. Helwig et al. \cite{Helwig+2015} elaborated on this condition-based monitoring system that is deployed in Germany. 

In line with the above-mentioned vision of the IoT, we assume that this system will be connected to the IoT in the future. In other words, each of the sensors and actuators involved will be directly connected to the Internet (IoT). One advantage of this will be the possibility of letting the condition-based monitoring systems deployed at multiple facilities or sites of one customer or a group of customers cooperate to the benefit of all of them. This might involve sharing their data to enhance the prediction performance of the ML models that are created and trained for different hydraulics systems. However, in the case that privacy concerns and regulations discourage or prohibit sharing raw data, they may use federated ML techniques, through which a number of systems deployed at various sites may cooperate in order to create a more capable joint ML model without sharing any raw data. The proposed approach in this work enables the modeling infrastructure for edge analytics and federated ML since it allows augmenting any arbitrary \textit{thing} with one or more data analytics component. Figure \ref{HydraulicsPredictiveMaintenance} illustrates the said hydraulics system that includes a primary working circuit and a secondary cooling and filtration circuit, as well as the predictive maintenance system for condition-based monitoring of the hydraulics system. The entire system is a CPS that is connected to the Internet (IoT). This use case scenario is an example of typical smart IoT services that can be modeled and their implementations can be automatically generated using the proposed approach.

\afterpage{
\vspace*{\fill}
\begin{figure}[!h]
	\centering 
	\includegraphics[angle=90,width=\linewidth]{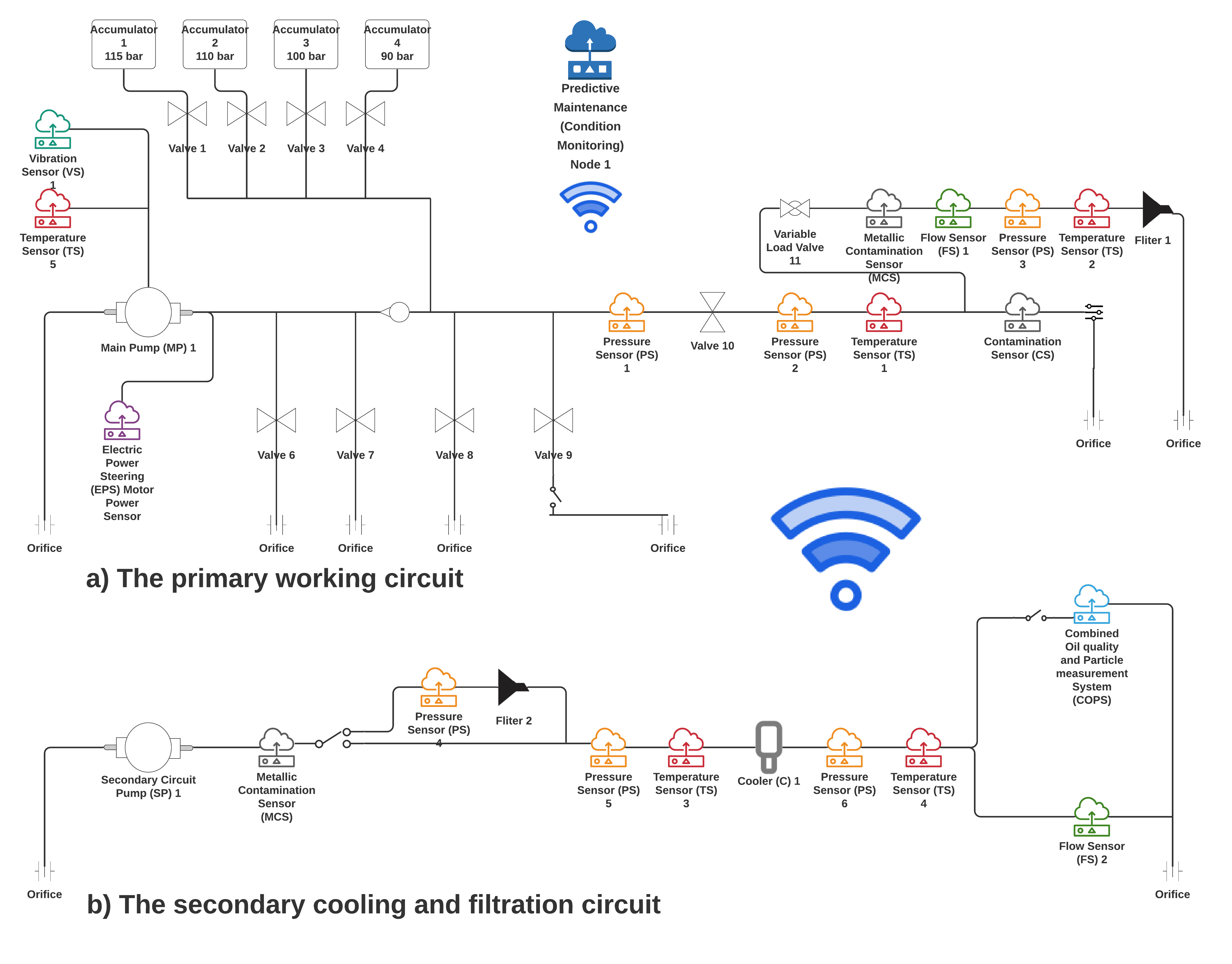}
	\caption{Predictive maintenance of a hydraulics system \cite{Helwig+2015}}
	\label{HydraulicsPredictiveMaintenance}
\end{figure}
\clearpage
\vspace*{\fill}
}

\subsection{Analytics Modeling}\label{background-analytics-modeling}

\textit{Analytics modeling} is a term that stands in contrast to \textit{analytics operations}. In fact, the core focus of the data analytics, also known as the Knowledge Discovery and Data Mining (KDD) community is on analytics modeling, which involves developing new algorithms, methods and techniques to manage and analyze data, e.g., for business intelligence, decision making support, optimization, predictive maintenance and so forth. One of the fields that has recently very much helped them in achieving their goal is ML (especially its sub-discipline \textit{deep learning}). Data scientists and ML engineers often practice analytics modeling. They usually offer the software that produces and trains DAML models, and are called (DAML) \textit{model producers}. However, in order to deploy and use DAML models in real-world systems, we also need data engineers, who together with software engineers, database engineers/designers and system engineers take other aspects, such as the performance and scalability of the entire system into account. The tasks of data engineers that mainly involve large-scale data analytics and processing (often referred to as \textit{big data analytics}) are grouped under the umbrella term analytics operations. Data engineers often provide the software that consume or use DAML models, thus called (DAML) \textit{model consumers}, also known as \textit{scoring engines} \cite{Pivarski+2016}. Note that in the stream processing (i.e., online learning) scenarios, where training the DAML model shall be an ongoing process that needs to be performed in a live manner, the boundaries between the mentioned groups of tasks may sometimes become blurred.

In the DAML community, the notion of \textit{models} is generally understood as the abstractions about the observed data that can help in understanding, analyzing and managing the data to generate value, e.g., to generate plausible instances of such data in order to make predictions. Leskovec et al. \cite{Leskovec+2014} referred to several common approaches to models in this community. For instance, one may define such a model as an underlying probability distribution, from which the observed data are presumably drawn. This is called the statistical approach. Alternatively, one may consider a model for a dataset to be a summarization or an approximation of its data instances. Further, some models represent a dataset by its most extreme examples. Those are called feature-based models. Finally, ML models that are currently widely used in analytics modeling - and are the main focus of this work - may come from diverse families, e.g., linear models, decision trees, ensemble models, such as random forests, kernel-based models, e.g., Support Vector Machine (SVM), Artificial Neural Networks (ANN) and Probabilistic Graphical Models (PGM) \cite{Bishop2006}. Deep ANNs with several hidden layers are currently widely used in the industry. Also, Bayesian Deep Learning \cite{WangYeung2020} is a promising approach for many industrial IoT/CPS use cases.

Furthermore, we need to clarify the terminology on model-based ML. Until recently (and even broadly today), model-based ML was (and is) understood as ML approaches that contrary to the so-called instance-based (also known as memory-based) ML approaches, they do not require storing any instances of the observed dataset that is used for training for the future uses. This means, the so-called model-based ML approaches, such as ANNs, have the ability to completely \textit{learn} the recognized patterns in the data and function independently of the observed data, once training is done. In contrast, instance-based approaches, e.g., SVMs require at least part of the observed dataset even after training in order to be able to work \cite{Bishop2006}. 

However, a nuanced notion of model-based ML, which is in line with the understanding of the SSE community from the term model-based, emerged with Infer.Net \cite{Bishop2013, InferNet}. According to this notion, which is also deployed here, model-based ML can be used with any ML model architecture, regardless of being instance-based or not.

\subsection{Software Modeling}\label{background-software-modeling}

In the SSE community, models are abstractions that describe the architecture of a software/system. Here, we are interested in software systems. Therefore, we concentrate on software models. Models can be at different levels of abstraction, thus having different degrees of details. Moreover, models may focus on different aspects of software systems. As long as a model can address the concerns of a stakeholder, it is interesting and relevant. A model instance shall conform to a meta-model, which specifies the syntax (and maybe also part of the semantics) of the corresponding modeling language. A modeling language might be general purpose, such as the Unified Modeling Language (UML) standard, or domain-specific, e.g., ThingML \cite{ThingML}. According to the ISO/IEC/IEEE 42010:2011 standard \cite{ISO-IEC-IEEE-42010-2011} for the architecture descriptions in systems and software engineering, an architecture description is made of one or often more architecture views. Several (software architecture) model instances may belong to one architecture view, which addresses one or several concerns of a stakeholder or a group of stakeholders. Based on the said standard, each architecture view is governed by one architecture viewpoint, which frames one or several concerns of a stakeholder or a group of stakeholders.

Further, if we consider the UML diagram notations, we observe that they can be categorized into two broad groups: (i) structural diagrams, e.g., the Class diagram, the Component diagram and the Object diagram; (ii) behavioral (including interaction) diagrams, e.g., the Activity diagram, the State machine diagram and the Use case diagram. The UML Activity diagram might be used for modeling the workflows (i.e., the flow of control) or data flows (i.e., the flow of data).

However, in this work, we are interested in Domain-Specific Modeling (DSM) with automated full code generation \cite{KellyTolvanen2008}, a MDSE approach that has been adopted both by the ThingML methodology \cite{ThingML, Harrand+2016} and ourselves \cite{ML-Quadrat, Moin+2018, Moin+2020}. Nevertheless, there exist other approaches to software modeling which either do not promise automated full code generation (e.g., they just generate a skeleton), or do not consider models as the central artifacts, i.e., they are not model-driven (model-based), but rather use models for specific tasks, such as designing, early prototyping and documentation. In this work, we are not interested in such approaches.

\section{Related Work}\label{related-work}
Raising the level of abstraction to hide the complexity, and providing partial or full automation - e.g., via model-to-code transformations for code generation out of software models, or via model-to-model transformations for transforming one model to another model conforming to a different meta-model - are two pillars of the MDSE paradigm, which treats software models as first-class citizens. Both of the said pillars have already been introduced to some extent in the field of DAML as well. Raising the level of abstraction has been practiced through libraries and frameworks with higher level APIs. For instance, TensorFlow \cite{Abadi+2015} offers a powerful API for deep learning using various advanced methods, while Keras \cite{Chollet+2015} provides yet a higher layer of abstraction, which supports both the APIs of TensorFlow and other deep learning frameworks, e.g., Theano \cite{Theano2016}. Moreover, DAML workflow designers, such as KNIME \cite{Berthold+2009} and RapidMiner \cite{RapidMiner}, and visualization toolkits, such as TensorBoard \cite{TensorBoard}, offer a graphical and abstract layer beyond the code. However, none of the mentioned approaches followed the systematic and holistic approach of the MDSE paradigm, where \textit{models} include the necessary information regarding the entire application, and model-to-code transformations are often capable of generating the software implementation out of them. The workflows in KNIME \cite{Berthold+2009} and RapidMiner \cite{RapidMiner} or the Computational Graphs (CG), also known as the Data-Flow Graphs (DFG) in TensorBoard \cite{TensorBoard}, which is the visualization toolkit for TensorFlow \cite{Abadi+2015}, never address any aspect or concern beyond DAML. Last but not least, some workflow designers, e.g., KNIME \cite{Berthold+2009} provide the partial code generation functionality for DAML. 

Furthermore, the idea of Model-Interchange Formats, such as Predictive Model Markup Language (PMML) \cite{PMML}, Portable Format for Analytics (PFA) \cite{PFA, Pivarski+2016} and Open Neural Network Exchange (ONNX) \cite{ONNX} is relevant to the principles and common practices of MDSE. PMML is an XML-based standard of the Data Mining Group (DMG) \cite{DMG}, which comes in the form of an XML-schema and is already supported by more than 30 vendors world wide. Also, PFA is an emerging standard of the DMG, which offers a much higher degree of flexibility and power compared to PMML. First, unlike PMML, that only supports a limited set of DAML models, PFA provides a DSL that enables the implementation of any DAML method. Second, with PFA one may model an entire workflow or pipeline, not just a single DAML model. In addition, ONNX supports building Artificial Neural Networks (ANN) models from various libraries and frameworks, e.g., TensorFlow \cite{Abadi+2015}, Keras \cite{Chollet+2015}, PyTorch \cite{Paszke+2017}, Scitkit-Learn \cite{Pedregosa+2011}, MXNET \cite{Chen+2015}, Caffe2 (which is now part of PyTorch \cite{Paszke+2017}), XLA (which is a domain-specific compiler for linear algebra that can accelerate TensorFlow models), Core ML (that allows integrating ML models into the iOS apps) and the Microsoft Cognitive Toolkit (previously known as CNTK) in an interoperable manner.

The second pillar of MDSE, namely automation has been also applied to the DAML field. Infer.Net \cite{InferNet, Bishop2013} proposed the idea of using ML models, specifically PGMs, as MDSE models, thus generating the entire software implementation automatically out of them. They only supported C\# for code generation. Although this approach to ML has so far been the most relevant approach to the MDSE paradigm, it has a major shortcoming for real-world IoT/CPS applications, where the expressiveness of PGMs and other ML models does not suffice to model the entire software system and generate the full source code out of the model instances.
 
Moreover, as set out in Section \ref{introduction}, ThingML \cite{Morin+2017, Harrand+2016, Fleurey+2011, ThingML} and HEADS \cite{Morin+2016, HEADS} supported the MDSE paradigm, specifically the DSM methodology \cite{KellyTolvanen2008} for full code generation in the IoT/CPS domain. While they mainly focused on the design-time of software systems, other approaches, such as $\mu$-Kevoree \cite{Fouquet+2012} concentrated on \textit{Models@Runtime}, thus fading out the borders between the design-time (modeling-time) and the runtime of IoT services. The major shortcoming of all of the said approaches is the lack of DAML support at the modeling level. In other words, the users of those DSMLs may not deploy the APIs of DAML libraries and frameworks in their software models. Hence, there is no seamless integration between the software models and the DAML models. In this work, we fill in this gap in the literature. We allow the DAML functionalities to be offered both by the cloud and by the edge devices. Therefore, our model-driven approach also supports edge analytics and federated learning by design.

The original idea of enhancing MDSE models for integrating ML models and software models has been proposed in our previous work, i.e., the position paper \cite{Moin+2018} and the poster/extended abstract \cite{Moin+2020}. In addition, Benoit et al. \cite{Benoit+2020} proposed a conceptual reference model for MDE of data-centric systems that helped in identifying different models, mainly ML models and software/system models, as well as their roles in the software/system life-cycle. In this manuscript, we formalize our prior work \cite{Moin+2018,Moin+2020}, realize its proof-of-concept and validate the underlying research hypotheses (see Section \ref{introduction}). 

Further, based on the Kevoree Modeling Framework (KMF) and $\mu$-Kevoree \cite{Fouquet+2012}, Hartmann et al. \cite{Hartmann+2017, Hartmann+2018, Hartmann+2019} proposed GreyCat \cite{GreyCat}, which integrated ML with software models in MDSE. Their idea and concepts were relevant to the work of Moin et al. \cite{Moin+2018, Moin+2020}. However, they only supported Java and Javascript/Typescript code generation, which was not sufficient for our purpose since we aim to cover code generation for the entire IoT systems that often consist of a range of heterogeneous IoT platforms, which may not be capable of running any Java Virtual Machine (JVM) at all, due to their resource constraints. Therefore, we build our approach on ThingML \cite{Morin+2017, Harrand+2016, Fleurey+2011, ThingML}. 

Finally, Table \ref{tab:related-work} compares the related work in the literature with the proposed approach. As we can see, the proposed approach, ML-Quadrat has all the benefits of the state of the art in MDE for the IoT (MDE4IoT), namely ThingML \cite{Morin+2017, Harrand+2016, Fleurey+2011, ThingML} and HEADS \cite{Morin+2016, HEADS}, but can also support DAML and integrate DAML models with the SE models.

\begin{table}[htbp]
	\caption{Related work in the literature compared to the proposed approach (ML-Quadrat)}
	\begin{center}
		\begin{tabular}{|p{1cm}|p{2.5cm}|p{1cm}|p{1cm}|p{1cm}|p{1cm}|}
			\hline	
			\textbf{Desc- ription} & \textbf{Work}  & \textbf{Full code gen.} & \textbf{DAML support} & \textbf{IoT / CPS domain} & \textbf{Model type} \\
			\hline
			ML libraries and frameworks & TensorFlow \cite{Abadi+2015}, Keras \cite{Chollet+2015}, Scitkit-Learn \cite{Pedregosa+2011}, etc. &   & \centering \checkmark &  & DAML models \\
			\hline
			DAML workflow designers & KNIME \cite{Berthold+2009}, RapidMiner \cite{RapidMiner}, etc. &  & \centering \checkmark &  & DAML models \\
			\hline
			Model Interchange Formats (MIF) & PMML \cite{PMML}, PFA \cite{PFA, Pivarski+2016}, ONNX \cite{ONNX} &   & \centering \checkmark &  & DAML models \\
			\hline
			\lq{}Model-based\rq{} ML & Infer.Net \cite{InferNet, Bishop2013}  & \centering \checkmark & \centering \checkmark &  & ML (PGM) \& SE models \\
			\hline
			MDE4 IoT & ThingML \cite{Morin+2017, Harrand+2016, Fleurey+2011, ThingML} and HEADS \cite{Morin+2016, HEADS} &   \centering \checkmark &  & \centering \checkmark  & SE models \\
			\hline
			Models@ Runtime & $\mu$-Kevoree \cite{Fouquet+2012}  &  \centering \checkmark &  & Limited  & SE models \\
			\hline
			Models@ Runtime + ML & GreyCat \cite{GreyCat}  &  \centering \checkmark & \centering \checkmark & Limited  & SE \& DAML models \\
			\hline
			MDE4 IoT + ML & ML-Quadrat \cite{ML-Quadrat}  &  \centering \checkmark & \centering \checkmark & \centering \checkmark  & SE \& DAML models \\
			\hline
		\end{tabular}
		\label{tab:related-work}
	\end{center}
\end{table}

\section{Proposed Approach}\label{proposed-approach}
In this section, we propose a novel approach to MDE for both analytics modeling (with a focus on ML) and software modeling, particularly for the IoT use case domain. In the following, we first illustrate the overall architecture of the proposed approach in Section \ref{proposed-approach-architecture}. Then, we formalize the proposed approach in Sections \ref{proposed-approach-analytics-models}, \ref{proposed-approach-mdse-models} and \ref{proposed-approach-enhanced-mdse-models}. 
As stated in Sections \ref{introduction} and \ref{related-work}, we extend the open-source ThingML project \cite{Morin+2017, Harrand+2016, Fleurey+2011, ThingML}, including the abstract syntax, i.e., the meta-model (grammar), the concrete syntax (model editors) and the semantics that are mostly realized in the model-to-code transformations, also known as code generators (\lq{}compilers\rq{}). The proposed approach and its implementation (see Section \ref{open-source-prototype}) are backward compatible, thus interoperable with the ThingML \cite{ThingML} (and HEADS \cite{HEADS}) models and code generators. 
In particular, we augment the meta-model (grammar) of the DSML of ThingML \cite{ThingML} with a new component, called Data Analytics (DA), which is responsible for enabling Data Analytics and Machine Learning (DAML) at the modeling level, such that practitioners using the DSML can obtain access to the APIs of the DAML libraries and frameworks (e.g., Scikit-Learn \cite{Pedregosa+2011} and Keras \cite{Chollet+2015}) in their software models at the design-time. To this aim, we also have to extend the action types of ThingML \cite{ThingML} (see Section \ref{proposed-approach-mdse-models}). Additionally, we extend the Java code generator of ThingML \cite{ThingML} to generate Python code as well. The Python code, which is seamlessly integrated with the Java code, is responsible for realizing the DAML functionalities, using the APIs of Scikit-Learn \cite{Pedregosa+2011} and Keras \cite{Chollet+2015} (the latter with the TensorFlow \cite{Abadi+2015} backend).

\subsection{Overall Architecture}\label{proposed-approach-architecture}
\begin{figure*}[htb!]
	\centerline{\includegraphics[width=\textwidth]{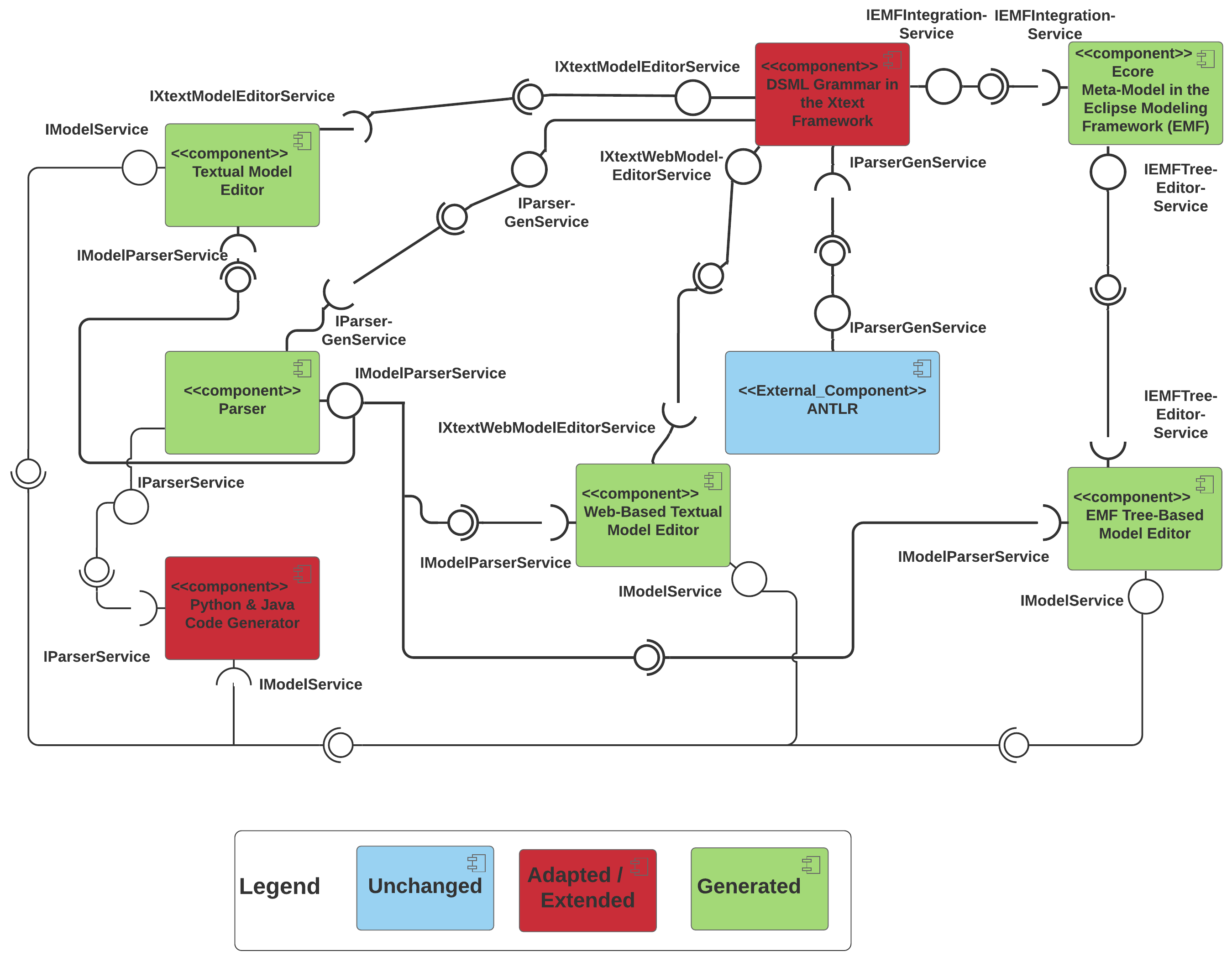}}
	\caption{The UML Component diagram illustrating the logical architecture view of the proposed approach.}
	\label{fig:uml-component}
\end{figure*}

The UML Component diagram that illustrates the logical view of a number of key functional software components is presented in Figure \ref{fig:uml-component}. Most of them were also present in the prior work, ThingML \cite{ThingML}. However, we adapted and extended them. According to the legend of the diagram, the unchanged, adapted/extended, and generated components are depicted in blue, red and green, respectively. Here, we skipped the rest of the code generators that are inherited from the ThingML \cite{ThingML} project, e.g., the C/C++ code generators.

Most importantly, we introduce the DAML concepts and functionalities into the DSML grammar in the Xtext framework as the main innovation concerning the meta-model (grammar). We discuss the new elements, such as the new action types in Sections \ref{proposed-approach-mdse-models} and \ref{open-source-prototype}. Many other components besides the modeling language grammar, shown in Figure \ref{fig:uml-component}, such as the Ecore meta-model, the model editors, namely, the textual model editor in the Eclipse Modeling Framework (EMF), the tree-based model editor in the EMF, and the web-based textual model editor (in-browser), as well as the parser are generated automatically out of this grammar. 

\begin{figure}[htbp]
	\centerline{\includegraphics[width=\textwidth]{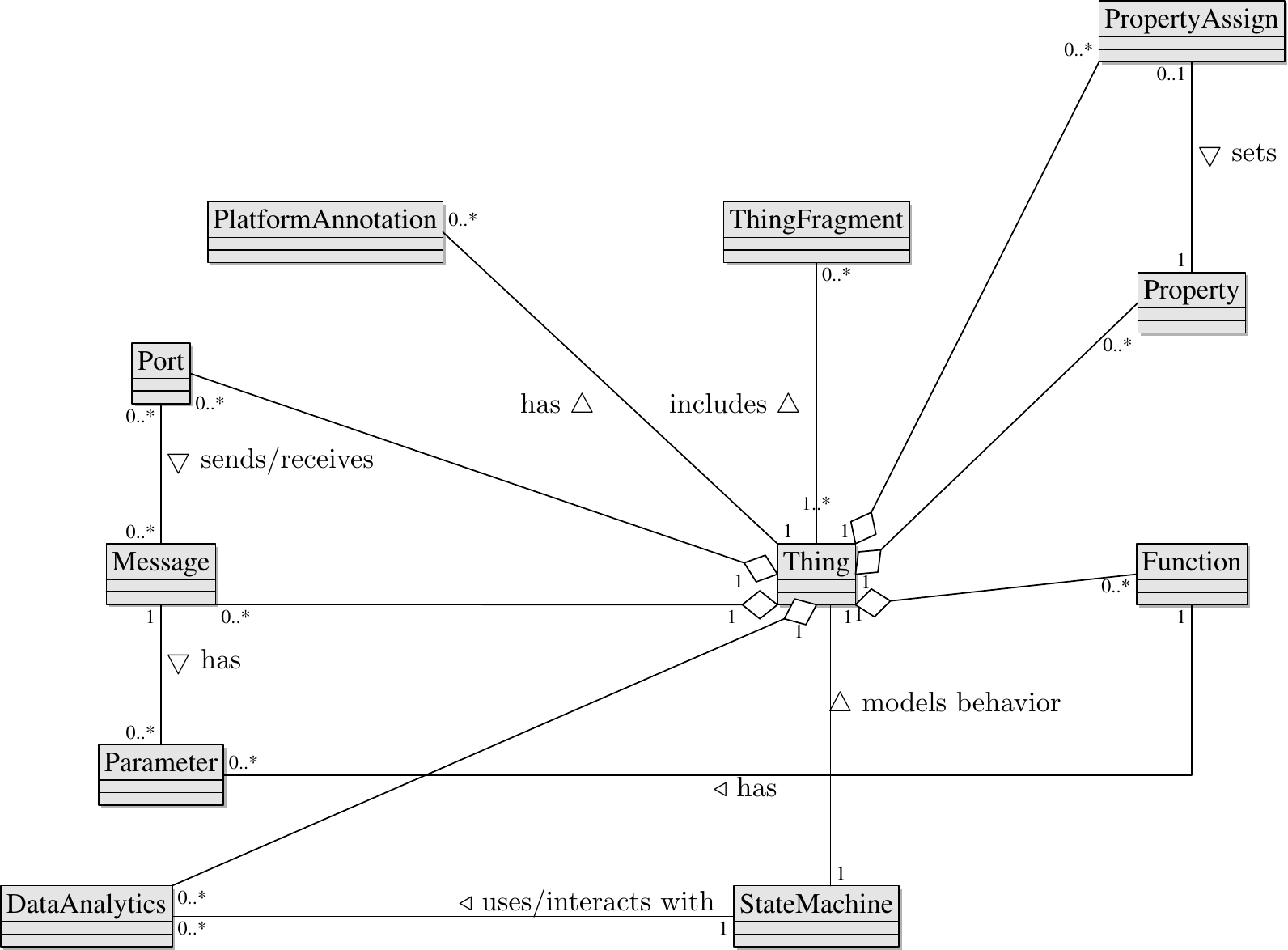}}
	\caption{The UML Class diagram showing part of the meta-model of the proposed DSML}
	\label{fig:MM}
\end{figure}

The UML Class diagram in Figure \ref{fig:MM} presents part of the abstract syntax (i.e., grammar or meta-model) of the proposed DSML\footnote{Note that almost every class, e.g., State Machine, Data Analytics, etc. is in practice associated with the Platform Annotation class. However, to prevent the figure from becoming cluttered, those associations are not shown here.}. Except for the Data Analytics class, the rest has been adopted from the prior work, ThingML \cite{ThingML}. Therefore, we allow each of the \textit{things} to optionally include one or more Data Analytics (DA) components that are in charge of carrying out DAML tasks, such as predictions. The focus of the DAML part is mainly on the ML methods and statistical inferences rather than simple analytics via some basic statistics or rule-based engines. Currently, we handle supervised and unsupervised ML.

\begin{figure*}[htb!]
	\centerline{\includegraphics[width=0.75\textwidth]{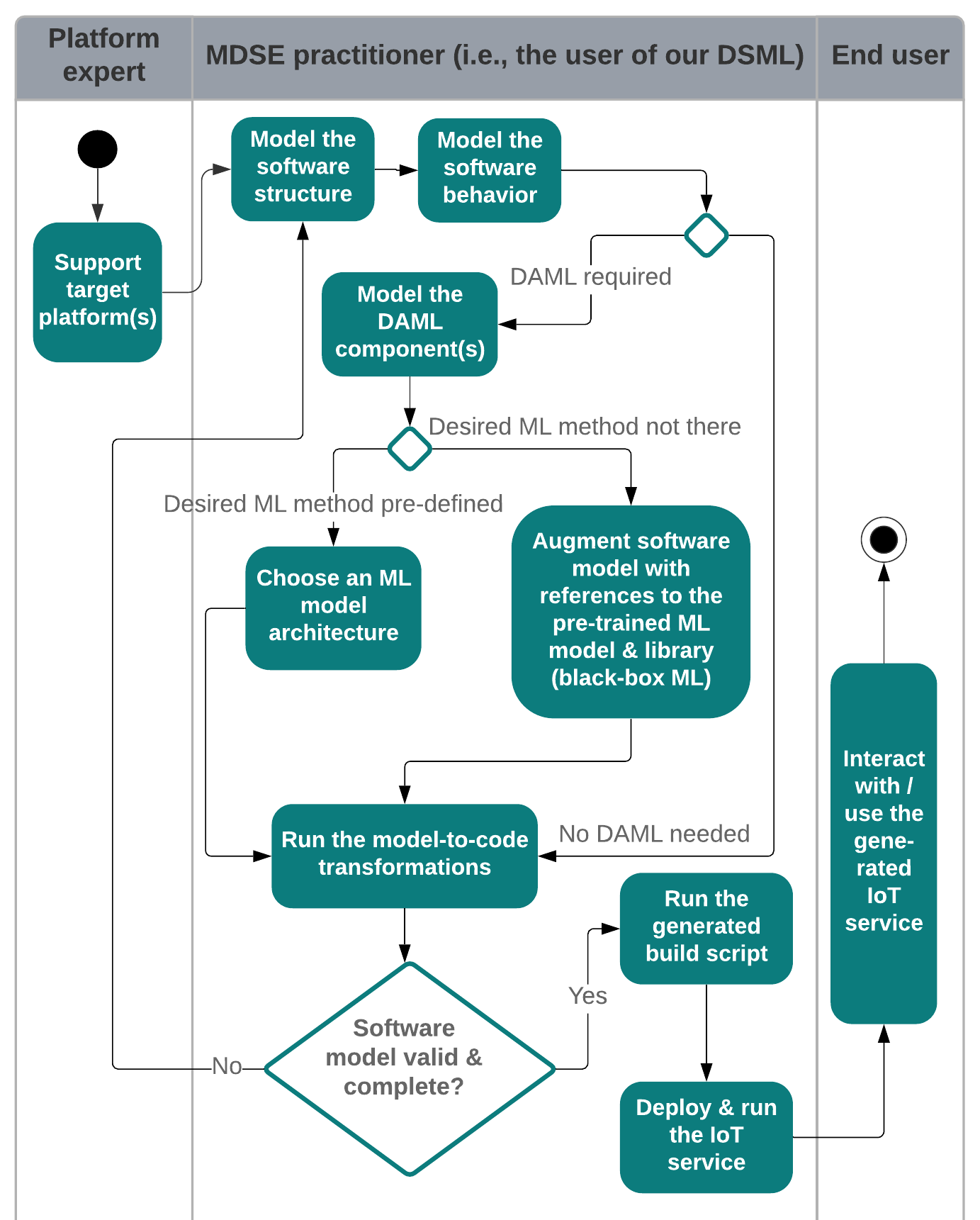}}
	\caption{The UML Activity diagram illustrating the usual workflow of using the proposed approach.}
	\label{fig:uml-activity}
\end{figure*}

Finally, Figure \ref{fig:uml-activity} depicts the UML Activity diagram that shows the usual workflow for deploying the proposed approach in the software development process of smart, data-driven IoT services. 

\subsection{Analytics Models (Focused on ML Models)}\label{proposed-approach-analytics-models}
We define an ML model, called $DM$ (the abbreviation of \textbf{D}ata \textbf{M}odel) used in analytics modeling as follows:

\begin{equation}
  \label{eq:dm}
  DM = (\upsilon, P, \Phi, H, I)
\end{equation}

Here, $\upsilon$ is an argument that indicates the structure or family type of the ML model $DM$, e.g., Decision Tree (DT), Probabilistic Graphical Model (PGM) or Multi-Layer Perceptron (MLP) Artificial Neural Network (ANN), P is a set which contains all of the \textit{parameters} of the model $DM$ with their respective values, $\Phi$ indicates the sequence of ML \textit{features} (i.e., ML \textit{attributes} and their values) with their respective data types, $H$ is the set of all \textit{hyperparameters}, e.g., the \textit{optimization or learning algorithm} $\zeta$ that shall be used to \textit{train} the model $DM$, the choice of the \textit{error/loss/cost/objective function} $e$, the \textit{batch size} $bs$, the number of \textit{epochs} $ne$, the \textit{learning rate} $lr$ if applicable, etc., and $I$ is the set of additional information or meta-data about the model and/or the data. $I$ might include the following items: (i) Whether the model is already trained, if applicable what the training stage is and when the time of the last training was; (ii) The paths or URIs/URLs of the dataset(s) used for training, validation and testing; (iii) Whether any of the data instances has a label (in that case the last item of the sequence of features $\Phi$ indicates the ML class labels and its data type\footnote{We also support array labels/outputs. In the future, we plan to support Sequence-to-Sequence models as well (see Section \ref{conclusion-futurework}).}); (iv) If the dataset is sequential, e.g., time series, so that the order of the data instances matter; (v) Whether the training is performed online, i.e., stream processing or offline, i.e., batch processing. In the former case, the dataset is virtually unbounded, whereas in the latter case, the dataset is bounded.

Analytics modeling involves designing the model $DM$, and then training it, which means using $\zeta$ and other hyperparameters in $H$ to fine-tune the values of the parameters in P, so that $DM$ can then make reasonable predictions $Y_{pred}$  for the previously unobserved data instances, say $X_{new}$, where the amount of the error/loss, $e$ for the prediction of $DM$ given the unobserved inputs, i.e., $pred(DM, X_{new})$ remains below a certain threshold $\varepsilon$:

\begin{equation}
  \label{eq:dm-training}
  DM=(\upsilon, P, \Phi, H, I), \quad train(DM) \rightarrow E[e(pred(DM, X_{new}))] < \varepsilon
\end{equation}

Here, $E$ is the expected value and $e$ is the error/loss, which might be defined according to various metrics, e.g., the \textit{Mean Absolute Error (MAE)}, also known as the \textit{L1-norm} for regression:

\begin{equation}
  \label{eq:dm-mae}
  e = \frac{1}{n}\displaystyle\sum_{i=1}^{n} \mid \hat{y}_i - y_i \mid
\end{equation}

In the equation above, $n$ is the number of data instances, $\hat{y}_i$ is the predicted numerical label by $DM$ for the $i_{th}$ data instance, and $y_i$ is the actual numerical label of this data instance.

As mentioned, the choice of the metric for $e$, e.g., MAE, is specified in the hyperparameters $H$. Moreover, hyperparameter tuning is an important part of the analytics modeling practices. Currently, this has to be done manually. In the future, we plan to support more Automated ML (AutoML) functionalities to offer automated hyperparameter tuning too (see Section \ref{conclusion-futurework}).

If the data instances are labeled, the task is a \textit{supervised} ML task, thus the prediction implies finding the correct class label for a new, previously unobserved data instance. However, if the data instances do not possess class labels, it is called an \textit{unsupervised} ML task. For instance, in the case of \textit{clustering}, which is an example for unsupervised learning, prediction refers to finding the right cluster for each new data instance. In many applications, only some instances may already have class labels and some or many of them may not have one. This latter case is called \textit{semi-supervised} learning. Further, a supervised ML task with numerical class labels is called \textit{regression}, whereas a supervised ML task with categorical class labels is known as \textit{classification}.

\subsection{Software Models (in domain-specific MDSE for the IoT)}\label{proposed-approach-mdse-models}
We define a software model, or more precisely a software architecture model instance, called $SM$ as shown in Equation \ref{eq:sm}, where $\Psi$ is the set of structural elements, and $B$ is the set of behavioral elements.

\begin{equation}
  \label{eq:sm}
  SM = (\Psi, B)
\end{equation}

However, since we are interested in domain-specific MDSE with automated full code generation, we augment the said software model formulation with a set of annotations, $A$ and a set of configurations, $C$, thus as defined in Equation \ref{eq:sm-dsm}.

\begin{equation}
  \label{eq:sm-dsm}
  SM = (A, \Psi, B, C)
\end{equation}

\paragraph{Annotations}
The Annotations ($A$) often help attach additional semantics to model instances. For example, one may specify which of the available library (API) choices for a certain task, such as ML methods, or the communication protocols shall be used for code generation. This means, if, for example, both Scikit-Learn and Keras offer a certain ML model/algorithm, which is desired, e.g., the MLP-ANN, one may choose through an annotation whether the APIs of Scikit-Learn or the APIs of Keras must be generated by the model-to-code transformation that generates Python code.

\paragraph{Structural elements}
The structural elements ($\Psi$) specify the \textit{static} aspect of the software system. In the IoT/CPS context (see the use cases in Section \ref{validation-case-study}), $\Psi$ consists of the \textit{things} $T$ (in the sense of IoT cloud and edge devices in a distributed system), and for each \textit{thing} $\tau_i \in T$, the ports $P_i$ for communication with other \textit{things} $\tau_j, j \neq i$, the messages $M_{p_i}$ associated to each port for message-passing, and the properties or local variables $\Gamma_i$. Each message $m_{p_{i_j}} \in M_{p_i}$ must have a direction (inbound/outbound) and may include one or more parameter(s) $par(m_{p_{i_j}}) \in Par(m_{p_{i_j}})$. Both the properties/variables $\gamma_{i_j} \in \Gamma_i$ and the message parameters $par(m_{p_{i_j}}) \in Par(m_{p_{i_j}})$ are \textit{typed}, e.g., integer, float/double, String, etc. How each of the mentioned types in the model instance shall be translated or mapped to the specific types of the target platforms for code generation, e.g., whether the type integer shall be mapped to short, int or long in Java, must be set through the annotations $a_i \in A$.

\paragraph{Behavioral elements}
The behavioral elements ($B$) specify the \textit{dynamic} aspect of the software system. We consider a Finite-State Machine (FSM) (also known as a finite-state automaton) model, called $FSM_i \equiv B_i$ for the behavior of each of the \textit{things} $\tau_i \in T$. We define the FSM model as follows:

\begin{equation}
  \label{eq:sm-fsm}
  FSM = (\Sigma, S, s_0, \delta, F, \Pi) 
\end{equation}

Here, $\Sigma$ is a set of inputs (explained below) which must be finite and non-empty by definition, $S$ is a set of states for the thing $\tau_i \in T$ which is also finite and non-empty, $s_0 \in S$ is an initial state that must be specified, $\delta : S \times \Sigma \rightarrow S$ is the state-transition function, $F \subseteq S$ is a (possibly empty) set of final states, and $\Pi$ is a set of actions (illustrated below). In this work, we assume the finite-state automaton to be deterministic, i.e., given an input and a particular state, there will be only one output state for the transition function $\delta$, not a set of states. 

Moreover, since we adopt the event-driven programming paradigm, which is a natural fit for reactive and interactive IoT systems, the inputs $\sigma_i \in \Sigma_i$ in $FSM_i \equiv B_i$ (i.e., the behavioral model of $\tau_i \in T$) are basically events, e.g., the incoming messages sent from other things $\tau_j \in T , j \neq i$ to $\tau_i$. However, the actions $\pi_i \in \Pi$ may be diverse actions, such as printing a text in the standard output, storing a message $m_{p_{i_j}}$ or one of the parameters of a message $par(m_{p_{i_j}})$ in a local variable (property) $\gamma_{i_j}$ of the thing, or sending a message from $\tau_i$ to another thing $\tau_k \in T,  k \neq i$. The new action types that we added to the existing DSML of ThingML \cite{ThingML} are the following ones for DAML: (i) \textit{DA\_Preprocess:} This action results in pre-processing the data and making them ready for training the ML model. (ii) \textit{DA\_Train:} This action leads to performing ML model training. (iii) \textit{DA\_Predict:} This action enables asking the ML model for prediction. (iv) \textit{DA\_Save:} This action supports appending the prediction of the ML model to the dataset that was used for training the ML model. Please note that the trained ML models that are resulted from the \textit{DA\_Train} action will be serialized and stored in any case regardless of the \textit{DA\_Save} action.

\paragraph{Configurations}
The configurations ($C$) include a set of instantiations of the things, which is analogous to object instantiation from the classes in the Object-Oriented Programming (OOP) paradigm. Also, it is at this place of the model instance where the desired connections between the ports of the instantiated things are set out. Last but not least, configurations may optionally also include annotations, e.g., specifying which model-to-code transformations shall be used for code generation, and/or which communication protocols shall be employed (e.g., MQTT, HTTP, CoAP). Hence, we define a configuration $C_i$ for $\tau_i \in T$ as follows:

\begin{equation}
  \label{eq:sm-conf}
  C_i = (A_{C_i}, \Theta, \Xi)
\end{equation}

In Equation \ref{eq:sm-conf}, $A_{C_i}$ is the set of annotations for the configuration, $\Theta$ is the set of instances of things and $\Xi$ is the set of connectors between the ports of two things. Each instance $\theta \in \Theta$ has an instance name and a type, i.e., the corresponding thing $\tau_i \in T$. Further, a connector $\xi \in \Xi$ has a starting point, i.e., a thing instance and its port $\theta_a.p_j$, as well as an end point, i.e., another thing instance and its port $\theta_b.p_k$.

Finally, in the adopted domain-specific MDSE methodology with full code generation in an automated manner (see \cite{KellyTolvanen2008, Harrand+2016}), the assumption is that the software model $SM$ contains sufficient amount of information (i.e., it is semantically \textit{complete}) and is syntactically correct (i.e., it is \textit{valid}) according to the meta-model or the context-free grammar of the modeling language, so that the model-to-code transformations can generate the entire implementation of the software for the respective target hardware and software platforms out of the model instance $SM$. Formally, this means:

\begin{equation}
  \label{eq:sm-codegen}
  \exists \Delta, \quad is\_valid(SM)\ \&\ is\_complete(SM) \rightarrow \Delta(SM) \equiv full\_source\_code
\end{equation}

Here, $\Delta$ is a model-to-code transformation, $is\_valid$ returns a Boolean value that is true if and only if the model instance is valid, and $is\_complete$ returns a Boolean value that is true if and only if the model instance is complete. The parser and the model editor that we inherited from the ThingML project \cite{ThingML} and extended in this work concerning the DAML functionalities, support the user of the DSML to design a valid and complete model instance that conforms to the meta-model (grammar) of the DSML. The user of the DSML receives the possible error messages, warnings and hints for each of the lines of the textual model instance if applicable.

\subsection{AI-Enhanced MDSE Models (for smart IoT services)}\label{proposed-approach-enhanced-mdse-models}
Recall that we define a software model as shown in Equation \ref{eq:sm-dsm}. However, this corresponds to the classic approach to software systems, which tend to exhibit a pre-defined/fixed, stationary or static structure and behavior. Many intelligent systems today, especially for the IoT/CPS use case scenarios, pose a degree of dynamicity, where their structure and/or behavior may change, based on the runtime situation, e.g., the data coming from the surrounding environment. Therefore, either their structure or their behavior, or maybe even both, may be affected by the AI components of the system over the time. The proposed approach in this manuscript deploys ML to let the software model become adaptable. In other words, we propose considering $\Psi$ and/or $B$ as functions of ML models. We call this AI/ML-enhanced software model, Smart Software Model (SSM), and formalize it in the following way:

\begin{equation}
  \label{eq:ai-enahnced-sm}
  SSM = (A, f_\Psi(DM_1), f_B(DM_2), C)
\end{equation}

Here, $DM_1$ and $DM_2$ are two ML models for learning and controlling the dynamicity of the structure and the behavior of the smart software model, respectively. Thus, the structure and the behavior turn into functions of these ML models.

In the present work, we remove $DM_1$ for simplicity, and only employ ML for the behavior of the software model. Thus, we consider the simplified form below for our current implementation and validation ($DM_2$ is renamed to $DM$):

\begin{equation}
  \label{eq:ai-enahnced-sm-behavior}
  SSM = (A, \Psi, f_B(DM), C)
\end{equation}

In Equation \ref{eq:ai-enahnced-sm-behavior}, $DM$ is considered to be the ML model as defined in Equation \ref{eq:dm}, $\Phi$ is the sequence of ML features (attributes) of the ML model, $<\phi_1, \phi_2, ...>$, and $\phi_i \in \Gamma$, i.e., the ML features are chosen from the local variables (properties) of the respective thing $\tau$. Note that if the data instances are labeled, i.e., we have a supervised ML task (either classification or regression), as mentioned in Section \ref{proposed-approach-analytics-models}, the last item of the sequence of ML features $\Phi$ is considered as the class label, which shall be predicted by the ML model for new data instances. In practice, the local variables (properties) $\gamma_i \in \Gamma$ may be used in order to store the incoming messages and/or their parameters, so that they can be employed as ML features. Also, they can be used for storing the prediction of the ML model, e.g., to be used in a message, or to trigger an action by the same or another \textit{thing}. 

\section{ML-Quadrat: Open-Source Prototype}\label{open-source-prototype}

In this section, we present our open-source prototype, called ML-Quadrat, which implements the proposed approach. This prototype is used for the case study that is illustrated in Section \ref{validation-case-study}. The source code, the documentation and a number of examples are available in our Github repository \cite{ML-Quadrat} under the terms of the Apache License Version 2.0. Our prototype is built on top of the ThingML project \cite{ThingML}, which is also based on the Eclipse Modeling Framework (EMF) and the Xtext framework.

Furthermore, we offer a web-based version of the prototype that is not included in the open-source distribution, but is available upon request for the reproducibility of the results of the empirical evaluation in Section \ref{validation-empirical-eval}. The web-based interface helps us conduct the experiments with the external evaluators as they do not need to install any software on their side, but simply use the web application in their web browsers.

In the following, we first illustrate the abstract syntax and the concrete syntax of the DSML in Sections \ref{open-source-prototype-abstract-syntax} and \ref{open-source-prototype-concrete-syntax}, respectively. Then, we explain the model-to-code transformations (code generators) that realize the semantics and generate the full source code out of the software model instances, in Section \ref{open-source-prototype-semantics-m2c-transformations}. Further, we elaborate on the DAML matters, specifically on the ML methods that are supported out-of-the-box in the DSML, as well as how to deploy them, in Section \ref{open-source-prototype-supported-ml-methods}. However, we also enable the practitioners (e.g., software developers, data scientists and ML experts) who use the proposed approach, to deploy any arbitrary ML method in the so-called \textit{Black-box ML mode}. This is explained in Section \ref{open-source-prototype-mixed-mode-bakcbox-ml}. Finally, in Section \ref{open-source-prototype-sample-iot-service} below, we demonstrate a sample IoT service, which is a basic client-server interaction (ping-pong) to highlight the advantages of our work compared to the prior work, ThingML \cite{ThingML}.  

\subsection{Abstract Syntax of the DSML}\label{open-source-prototype-abstract-syntax}
The abstract syntax of the proposed DSML is defined in its grammar that is implemented with the Xtext framework. This is available in the source code repository of the open-source project on Github \cite{ML-Quadrat}.\footnote{See \href{https://github.com/arminmoin/ML-Quadrat/blob/master/ML2/language/thingml/src/org/thingml/xtext/ThingML.xtext}{https://github.com/arminmoin/ML-Quadrat/blob/master/ML2/ language/thingml/src/org/thingml/xtext/ThingML.xtext}}. The Ecore meta-model of the DSML is generated automatically out of the Xtext grammar. As mentioned in Section \ref{proposed-approach-architecture}, Figure \ref{fig:MM} depicts part of the meta-model of the DSML using a UML Class diagram.

As stated in Section \ref{proposed-approach}, the \textit{Data Analytics} class that is shown in Figure \ref{fig:MM}, which realizes $DM$ in Equation \ref{eq:dm} (see Section \ref{proposed-approach-analytics-models}), was not present in the prior work, ThingML \cite{ThingML}. This is explained in Section \ref{open-source-prototype-supported-ml-methods} and via the sample IoT service that is illustrated in Section \ref{open-source-prototype-sample-iot-service}. However, the rest has been adopted from the ThingML project \cite{ThingML} and partially extended to make it compatible with the proposed approach. 

Most importantly, the imperative \textit{action} language of ThingML \cite{ThingML} that supports event-driven programming on the state machines, which realize the behavioral models (i.e., $B$ in Section \ref{proposed-approach-mdse-models}) of \textit{things}, is extended. Using this action language, one may specify which actions (see $\Pi$ in Section \ref{proposed-approach-mdse-models}) must be taken upon the occurrence of a particular event, such as upon the receipt of a certain message type on a specific port of a \textit{thing}. For example, a state transition might happen due to the event. Also, various types of actions, such as conditional actions, loop actions, print actions, etc. were possible with the prior work. However, we introduced the new action types that were named in Section \ref{proposed-approach-mdse-models} in order to enable the DAML functionalities, namely creating and running the data pre-processing pipeline (i.e., \textit{DA\_Preprocess}), conducting ML model training (i.e., \textit{DA\_Train}), making predictions using the trained ML models (i.e., \textit{DA\_Predict}), and optionally saving the predictions in the dataset (i.e., \textit{DA\_Save}).\footnote{Trivially, DA\_Preprocess and DA\_Train are skipped in the case of a pre-trained ML model (see Section \ref{open-source-prototype-mixed-mode-bakcbox-ml}).} Section \ref{open-source-prototype-sample-iot-service} illustrates this using a simple example. 

Additionally, \textit{Thing} (\textit{Thing Fragment}), \textit{Platform Annotation}, \textit{Port}, \textit{Message}, \textit{Parameter} and \textit{Property} (see Figure \ref{fig:MM}) realize $\tau \in T$, $a \in A$, $p \in P$, $m \in M$, $par(m) \in Par(m)$ and $\gamma \in \Gamma$, respectively, that are mentioned in Section \ref{proposed-approach-mdse-models}. Last but not least, other elements, such as \textit{Function} and \textit{Property Assign}, as well as those which are not shown in Figure \ref{fig:MM}, fall outside of the scope of the focus of this work, thus can be found in the related work, for example, \cite{Morin+2017, Harrand+2016, Fleurey+2011, ThingML}.

\subsection{Concrete Syntax and Model Editors}\label{open-source-prototype-concrete-syntax}
\begin{figure}[htbp]
	\centerline{\includegraphics[width=\textwidth]{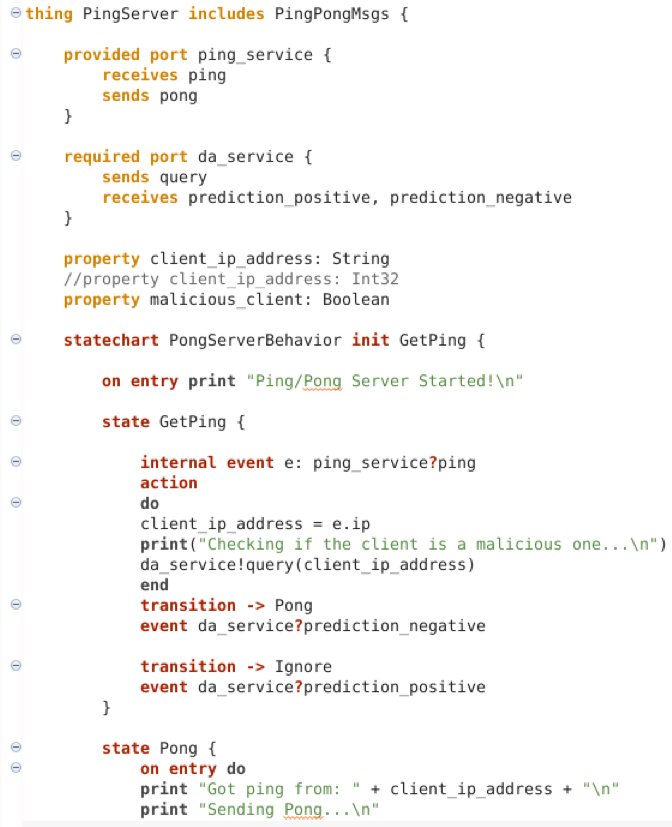}}
	\caption{The textual model editor, showing part of a sample model for the PingPong example (see Section \ref{open-source-prototype-sample-iot-service}).}
	\label{fig:texual-model-editor} 
\end{figure}

\begin{figure}[htbp]
	\centerline{\includegraphics[width=\textwidth]{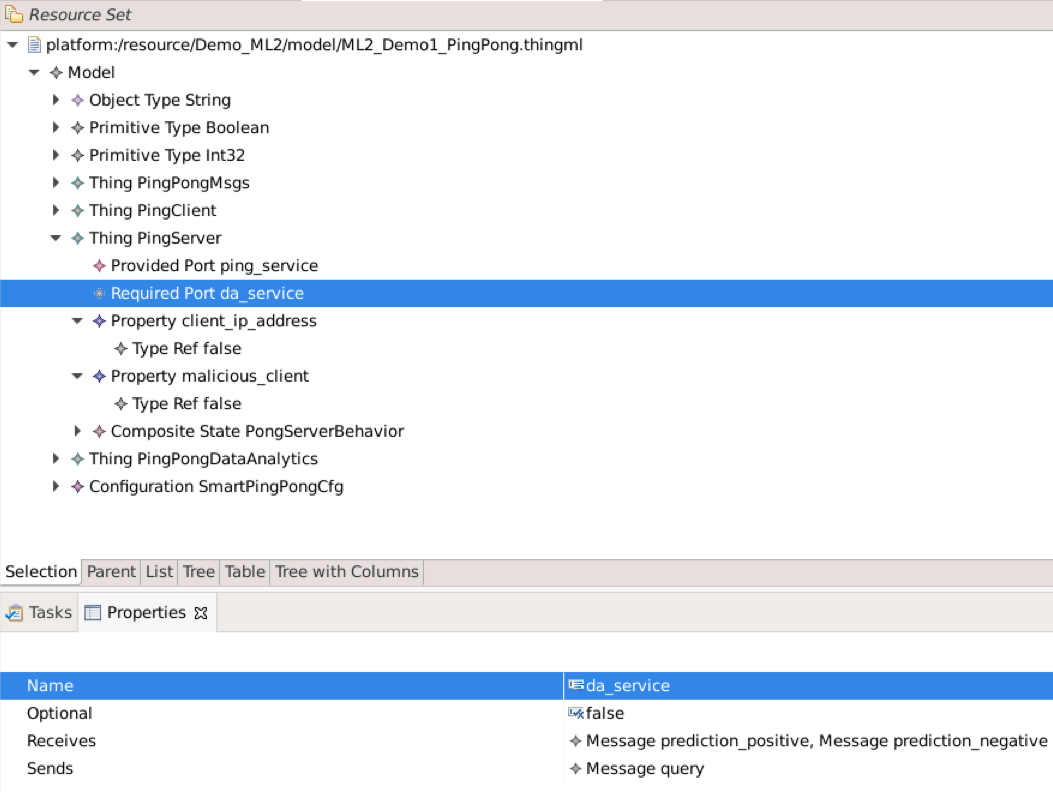}}
	\caption{The graphical, EMF tree-based model editor, showing part of a sample model for the PingPong example (see Section \ref{open-source-prototype-sample-iot-service}).}
	\label{fig:GraphicalModelEditor} 
\end{figure}

\begin{figure}[htbp]
	\centerline{\includegraphics[width=1.0\textwidth]{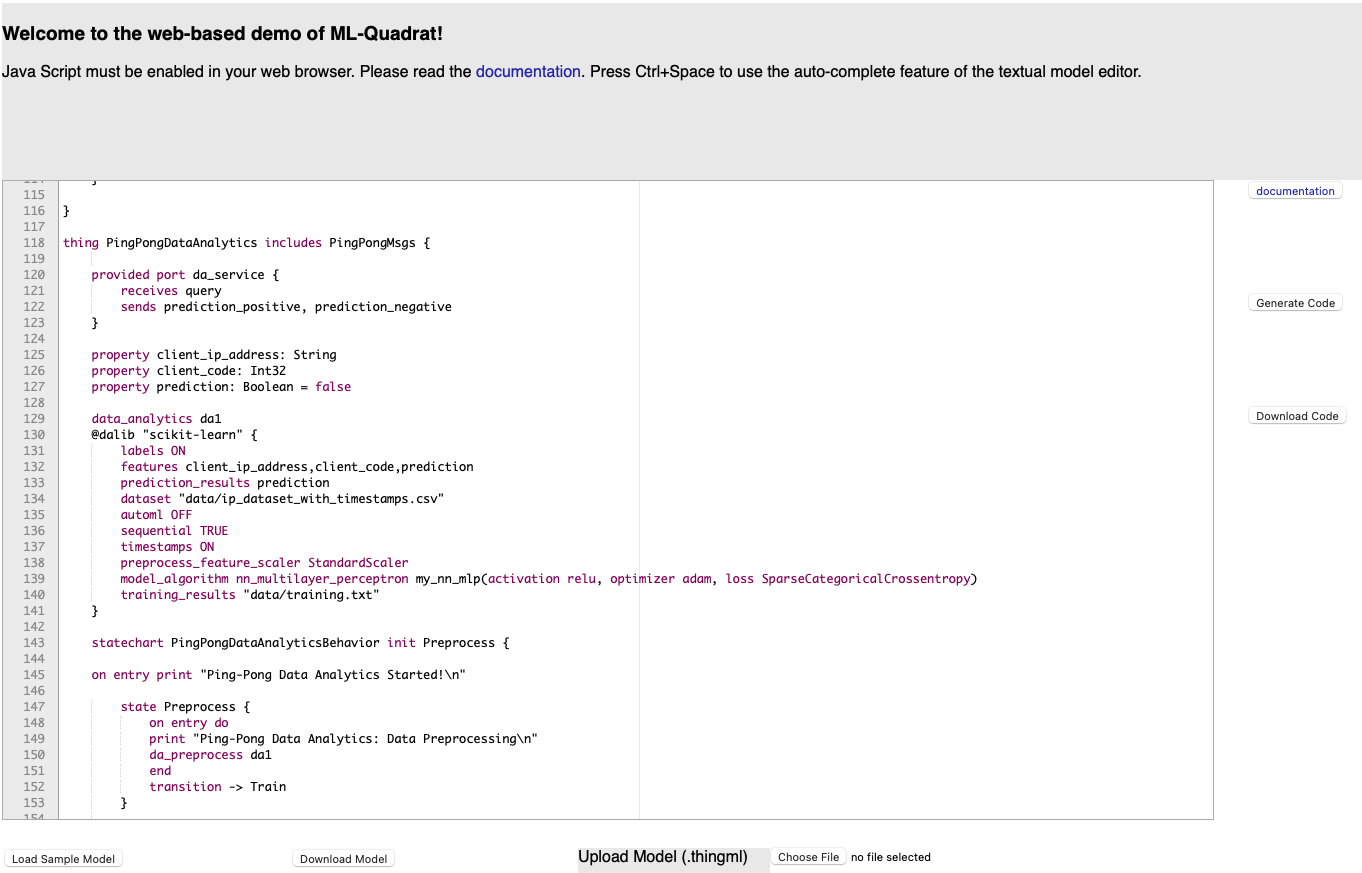}}
	\caption{The web-based prototype, showing part of a sample model for the PingPong example (see Section \ref{open-source-prototype-sample-iot-service}).}
	\label{fig:WebBasedPrototype} 
\end{figure}

We provide three model editors. First, a model editor, based on Xtext, is available in the EMF. This posses a textual concrete syntax, as well as the syntax highlighting and auto-complete features, and can give a number of hints and tips to help the practitioner (i.e., the user of the modeling tool) in designing a valid and complete model instance, out of which code generation for a working IoT service with the desired functionality is feasible. Figure \ref{fig:texual-model-editor} shows this model editor. Second, we offer a tree-based (form-based) model editor through the EMF. This is automatically generated in the EMF out of the Ecore meta-model of the DSML, which is itself generated automatically out of the Xtext grammar of the DSML. The tree-based model editor is demonstrated in Figure \ref{fig:GraphicalModelEditor}. While the textual version might be more suitable for developers, the tree-based editor might suite domain experts of the target IoT domains without software development skills well, so that they can modify certain properties of the software model instances, e.g., for the maintenance, upon possible future changes in the requirements. Last but not least, we develop a web-based prototype using the Java Servlets technology and the Xtext web integration. This web application offers a textual model editor with the auto-complete feature and some basic syntax highlighting. This is depicted in Figure \ref{fig:WebBasedPrototype}.

\subsection{Semantics and Model-to-Code Transformations}\label{open-source-prototype-semantics-m2c-transformations}
Part of the semantics of the DSML are included in the model-to-code transformations (i.e., $\Delta$ in Section \ref{proposed-approach-mdse-models}), also known as code generators or \textit{compilers}, and the associated constraint-checking mechanisms, which shall execute before the code generation. In addition, another part of the semantics is integrated into the grammar or meta-model, to enable type-checking and enforcing certain constraints at the design-time through the model editors (i.e., before executing the code generators). Furthermore, a number of annotations (i.e., $A$ in Section \ref{proposed-approach-mdse-models}), e.g., concerning the datatype mappings on specific target platforms, the choice of specific libraries for DAML, particular communication protocols, and model-to-code transformations are allowed on the modeling layer.

The proposed approach supports code generation in Python and Java. The Python code is responsible for the DAML functionalities of the target IoT services, and supports the APIs of Scikit-Learn \cite{Pedregosa+2011} and Keras \cite{Chollet+2015} with the TensorFlow \cite{Abadi+2015} backend. The model-to-code transformations are implemented in Java and Xtend (which is a modern variant of Java). They can be found in our Github repository \cite{ML-Quadrat}. \footnote{See \href{https://github.com/arminmoin/ML-Quadrat/tree/master/ML2/compilers/python_java}{https://github.com/arminmoin/ML-Quadrat/tree/master/ML2/ compilers/python\_java}}

\subsection{Supported ML Methods and Techniques}\label{open-source-prototype-supported-ml-methods}
The proposed approach allows each \textit{thing} to possess one or more components for DAML. Thus, it supports not only analytics in the cloud, but also edge analytics. Unlike the behavioral component of \textit{things}, i.e., the state machine (statechart), the DAML component, called Data Analytics (DA) is not mandatory. To exhibit DAML capabilities, a \textit{thing} has to include a \textit{data analytics} section in its model. This component that realizes $DM$ in Equation \ref{eq:dm}, might affect the behavior of the \textit{thing}, modeled via the corresponding state machine. As mentioned before, this corresponds to $f_B(DM)$ in Equation \ref{eq:ai-enahnced-sm-behavior}. In other words, the behavior of the thing becomes a function of the DAML model. Hence, if a \textit{thing} has a data analytics part, this part shall emerge before the state machine section in the textual model instance, so that the actions specified in the state machine may use and refer to the data analytics component.

Below, we list and briefly explain the possible parameters and options in the said data analytics section of the ML-enhanced software model instances that conform to the meta-model (grammar) of the proposed DSML (see Figures \ref{lst:SmartPingPong1} and \ref{lst:SmartPingPong2}):

\begin{enumerate}
	\item \textbf{Data\_analytics}: This parameter determines the name of the DAML component, e.g., da\_1.
	\item \textbf{Dalib}: The optional \textit{@dalib} annotation specifies the name of the library or framework which must be used for DAML. If this is absent, or it is set to \textit{auto}, or the desired ML method is not implemented in the selected library, the tool will try to automatically select the best option in the Automated ML (AutoML) mode (i.e., if AutoML is ON, see below).
	\item \textbf{Labels}: This is a binary parameter. If it is ON, it implies that the ML task is supervised. Hence, the last item on the list of features (see below) will be considered as the label. If the data type of that item, defined as the data type of the corresponding property (local variable) of the thing is numeric, e.g., Integer or Float/Double, then the ML task is a regression task. Otherwise, it is a classification task. Furthermore, if the parameter is set to OFF, then the task is unsupervised, e.g., clustering. This parameter also partially realizes $I$ as referred to in Section \ref{proposed-approach-analytics-models}.
	\item \textbf{Features}: This is a list of the properties (local variables) of the thing which shall be considered as the ML features (attributes). The local variables might include the messages or parameters of the messages that shall be received from other \textit{things}. As stated above, these are all considered as ML features only if Labels is OFF. In the case that Labels is ON, then the last item is not considered as a feature, but rather as the label (i.e., the class label for classification, or the target value for regression). This parameter realizes $\Phi$ as introduced in Section \ref{proposed-approach-analytics-models}. Simultaneously, the features are properties (local variables) of the corresponding \textit{thing}, thus also partially realizing $\gamma \in \Gamma$ in Section \ref{proposed-approach-mdse-models}.
	\item \textbf{Prediction\_results}: This parameter determines the property (local variable) of the \textit{thing} in which the prediction result, i.e., the output of the ML model prediction must be stored. Note that the properties were denoted by $\gamma \in \Gamma$ in Section \ref{proposed-approach-mdse-models}. The value of this property can be then later used in the \textit{actions} of the state machine, in order to let the ML model affect the behavior of the \textit{thing}.
	\item \textbf{Dataset}: The path of the dataset on the file system that shall be used for training the ML model. This must be a CSV (Comma-Separated Values) file without a header line.
	\item \textbf{AutoML}: This is a binary parameter indicating whether the AutoML mode must be used. If set to ON, a number of AutoML functionalities will be supported that can assist the practitioner, especially the novice users in the DAML field. By default, this is set to OFF.
	\item \textbf{Sequential}: This is a Boolean parameter that indicates whether the input data are sequential, e.g., time series, where the order of data instances matter. In this case, shuffling and cross-validation must be avoided. This parameter partially realizes $I$ as referred to in Section \ref{proposed-approach-analytics-models}.
	\item \textbf{Timestamps}: This binary parameter states if the data instances have timestamps or not. If this is ON, it has at least two implications. First, if new messages or parameters shall be appended to the dataset (using the DA\_Save action), timestamps will be automatically added by the tool. Second, the DAML method will be informed that the first column in the dataset, i.e., the CSV file, must be considered as the timestamp. The expected format is \textit{dd-mm-yyyy HH:MM:SS}, e.g., \textit{17-03-2021 22:49:06} for \textit{March 17, 2021 at 10:49:06 pm}. Obviously, if the timestamps parameter is ON, it is very likely that we are dealing with time series, i.e., sequential data\footnote{Note that the reverse does not always hold, as e.g., DNA data are sequential, but not time series data.}. Therefore, if the sequential parameter is not specified, the AutoML service of the tool, if it is set to ON, will automatically set the sequential parameter to True. However, if the user explicitly states that sequential is False, then the decision will not be overridden. The timestamps parameter also partially realizes $I$ as referred to in Section \ref{proposed-approach-analytics-models}.
	\item \textbf{Preprocess\_feature\_scaling}: This parameter specifies the feature scaling technique that must be used in the data preparation (pre-processing) pipeline. If it is not mentioned, in the case that AutoML is ON, then the best choice of scaling for the respective ML model/algorithm (see below) will be selected. For instance, for the higher performance of Artificial Neural Networks (ANNs), having numerical data that possess a relatively similar scale is an extremely important factor. Thus, for example, standardization (also known as the Z-Score normalization) is automatically set in the AutoML mode. This parameter partially realizes $H$ as set out in Section \ref{proposed-approach-analytics-models}.
	\item \textbf{ML Model/Algorithm}: Here, one can specify the particular ML method, including the ML model architecture (family) that must be deployed, e.g., Multi-Layer Perceptron (MLP) ANN, Decision Tree, etc. Additionally, the hyperparameters, e.g., the choice of the error/loss function ($e$), the learning/optimization algorithm ($\zeta$), the learning rate ($lr$), etc. might be given in parenthesis. Each family of ML models may have a different set of possible hyperparameters. The auto-complete feature (usually activated by pressing the Control and Space keys together for the textual model editors) helps in finding the possible options. Further, the documentation of the prototype, as well as the API documentations of the target frameworks and libraries (e.g., Scikit-Learn) must be studied. Also, a number of exception handling and logging mechanisms are available to support the user of the tool. This parameter realizes $\upsilon$, as well as $H$ in Section \ref{proposed-approach-analytics-models}. The parameters of the ML model (i.e., P in Section \ref{proposed-approach-analytics-models}) are controlled by the hyperparameters ($H$) during the learning process.
	\item \textbf{Training Results}: This is the path of the text file in which the log of ML model trainings shall be stored. The log includes information about the time of each training and the chosen ML model/algorithm. This parameter also partially realizes $I$ mentioned in Section \ref{proposed-approach-analytics-models}.
\end{enumerate}

We can see how the above-mentioned parameters are used in practice in the basic example provided in Section \ref{open-source-prototype-sample-iot-service} below.

Currently, the following ML models and algorithms are supported for supervised ML (i.e., for labeled data) out-of-the-box: (i) Linear Regression, (ii) Logistic Regression for linear classification, (iii) Na\"ive Bayes (the Gaussian, Multinomial, Complement, Bernoulli and Categorical variants), (iv) Decision Tree (both Regressor and Classifier), (v) Random Forest (both Regressor and Classifier), (vi) the Multilayer Perceptron (MLP) ANN. The APIs of Scikit-Learn are used for the items (i) to (v). However, for the MLP ANN, i.e., (vi) both Scikit-Learn and Keras are supported. By default Keras will be used for this family of ML models. However, the user may explicitly set the library for DAML to Scikit-Learn to override this recommended setting. This is possible through the annotation \textit{dalib} at the \textit{data\_analytics} section of the model instance. Moreover, a number of other techniques, e.g., for data preparation, specifically standardization or normalization of the numerical features using various methods are provided.

Moreover, the unsupervised ML methods that are also pre-defined, thus supported out-of-the-box are as follows: (i) K-Means, (ii) Mini-Batch K-Means, (iii) DB-SCAN, (iv) Spectral Clustering and (v) Gaussian Mixture Model. The APIs of the Scikit-Learn library are used for enabling them.

If the desired ML model, algorithm or technique is not pre-defined, one may either extend the open-source prototype (see the online documentation on Github \cite{ML-Quadrat}), or use the so-called \textit{Black-box} ML mode (also known as the \textit{hybrid/mixed} MDSE/non-MDSE mode) as described in Section \ref{open-source-prototype-mixed-mode-bakcbox-ml} below. In the latter case, one can bring any arbitrary pre-trained ML model and \textit{connect} it to the MDSE model.

\subsection{The Black-box ML (Hybrid/Mixed MDSE/Non-MDSE) Mode}\label{open-source-prototype-mixed-mode-bakcbox-ml}
Suppose that one does not want to use an existing ML method which is already available in our prototype, or has already an existing, pre-trained ML model that they want to deploy. In this case, the \textit{Black-box ML} mode, also called the \textit{hybrid} or \textit{mixed} MDSE/Non-MDSE mode shall be preferred. The drawback here is that the software model will not have any clue about the deployed ML method. Therefore, the ML model seems to the software model as a black-box. However, the advantage is that the user of the DSML will achieve a much higher degree of flexibility concerning ML. Hence, they may, in principle, introduce any pre-trained ML model with any arbitrary architecture and trained with any learning algorithm, and \textit{connect} or \textit{plug} it into the software model.

This can be done by using a parameter, called \textit{blackbox\_ml} and setting its Boolean value to true. In this case, using the \textit{model\_algorithm} and the \textit{training\_results} parameters will not be allowed in the data analytics section of the model instance as no training is required by the AI-enhanced MDSE model. The pre-trained ML model has to be stored in a separate directory. The path of this directory must be given through a parameter, called \textit{blackbox\_ml\_model} in the data analytics section of the model instance. The pre-trained ML model might have been trained with or without the proposed approach. Moreover, the ML method which is imported from the corresponding DAML library must be specified using a parameter, called \textit{blackbox\_import\_algorithm}.

\subsection{Sample IoT Service}\label{open-source-prototype-sample-iot-service}
In this section, we illustrate an example from the ThingML project \cite{ThingML}, and elaborate on the shortcomings of ThingML \cite{ThingML} by showing our extended (\textit{smart}) version of this example. Moreover, this sample IoT service was among the use cases which we originally used to create our DSML and modeling tool. However, the use cases that are provided in the case study in Section \ref{validation-case-study} are deployed for validating the proposed approach.

\paragraph{Ping-Pong}
This example originally came from the ThingML project \cite{ThingML}. In a distributed system, there exist two nodes, called \textit{things}, that are connected to the IoT: (i) the ping client and (ii) the pong server. The \textit{things} are involved in a basic client-server interaction, where the server simply waits for incoming ping messages from the client. As soon as a ping message arrives, the server responds with a pong message. 

\paragraph{Smart Ping-Pong}
We argue that in a real-world scenario with an enormous number of clients, which may send a ping message to the server, the example above can be enhanced via ML, in order to prevent the so-called Distributed Denial of Service (DDoS) attacks. Hence, we introduce a new \textit{thing} that is responsible for DAML, in order to predict if a client is prone to be an attacker or not. Upon receiving a ping message, the server consults this new \textit{thing}, which might even be a \textit{thing fragment} for the server, to see if the ping message shall be responded to with a pong or it would be safer to ignore the request, and perhaps even put the client in a blacklist for a certain period of time. Note that this was not possible using the ThingML DSML \cite{ThingML}, whereas our extended version supports DAML at the modeling level. Using the proposed DSML, one may enhance the model instance to become capable of DAML.

\begin{figure}[htbp]
	\centerline{\includegraphics[width=\textwidth]{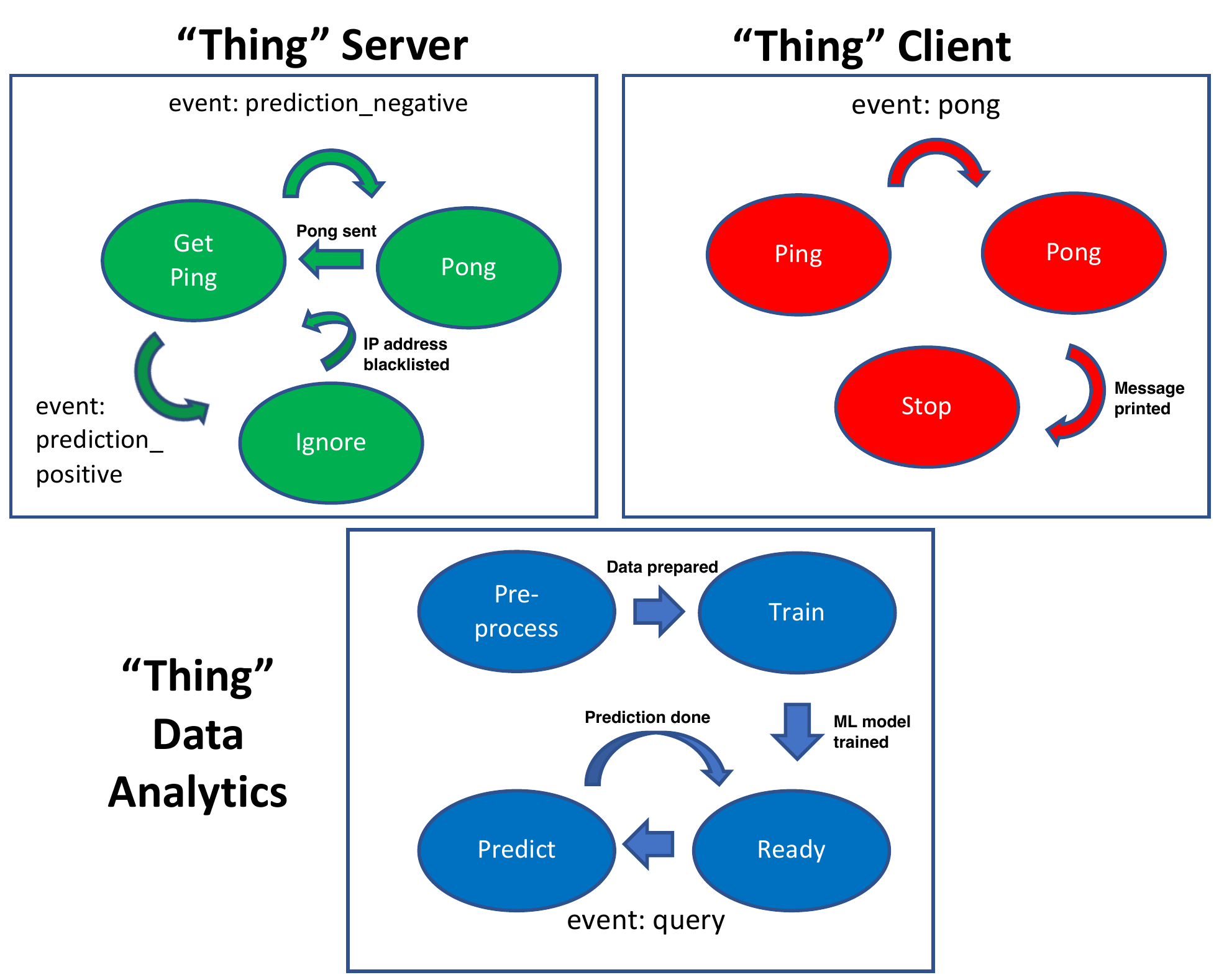}}
	\caption{The state machines modeling the behaviors of the three \textit{things} of the smart ping-pong example}
	\label{fig:SmartPingPongFSMs} 
\end{figure}

Figure \ref{fig:SmartPingPongFSMs} depicts the state machines that model the behaviors of the ping client, the pong server and the data analytics server.

Below, we demonstrate part of the model instance for the smart ping-pong example (see Figures \ref{lst:SmartPingPong1} and \ref{lst:SmartPingPong2}. The full model instance may be found in our Github repository\footnote{See \href{https://github.com/arminmoin/ML-Quadrat/blob/master/ML2/org.thingml.samples/src/main/thingml/ML2_Demo_PingPong.thingml}{https://github.com/arminmoin/ML-Quadrat/blob/master/ML2/org.thingml.samples/src/main/thingml/ ML2\_Demo\_PingPong.thingml}}.

\begin{figure}[!htbp]
\setlength{\fboxsep}{0pt}%
\setlength{\fboxrule}{0pt}%
\centering
  \lstset{language=Java, showspaces=false,
    showstringspaces=false, tabsize=2, breaklines=true,
    xleftmargin=5.0ex,
	basicstyle=\color{blue}\ttfamily\footnotesize\bfseries,
	commentstyle=\color{black}\ttfamily\footnotesize,
	frame=tb,
	linewidth=1.0\textwidth
}
\begin{lstlisting}
/* This is a part of the model instance. The full model instance is available in the Git repository on Github. */

thing PingPongDataAnalytics includes PingPongMsgs {
/* The messages are not shown here, but defined in a thing fragment, called PingPongMsgs. */

 provided port da_service { /* This port communicates with the da_service port of pingServer. */
  receives query /* This port may receive a query message from pingServer. */
  sends prediction_positive, prediction_negative /* This port may send a response to pingServer. The response might be positive, i.e., malicious prediction or negative, i.e., non-malicious prediction. */
 }
	
 /* The properties are the local variables of the thing. */
 property client_ip_address: String /* The IP address of pingClient is a String. */
 
 property client_code: Int32 /* This is just a secret integer code that is shared between pingClient and pingServer or alternatively a serial ID number for the ping message. */
 
 property prediction: Boolean = false /* This Boolean property shall store the prediction of the DAML model and is initialized as false here. This mean, by default, the client is non-malicious. */
	
 data_analytics da1 /* Please see Section 5.4. */
 @dalib "scikit-learn" {
    labels ON
    features client_ip_address,client_code,prediction 
    prediction_results prediction 
    dataset "data/ip_dataset.csv"
    automl OFF 
    sequential TRUE
    timestamps OFF
    preprocess_feature_scaler StandardScaler
    model_algorithm nn_multilayer_perceptron my_nn_mlp
    (activation relu, optimizer adam, loss SparseCategoricalCrossentropy)
    training_results "data/training.txt"
}

 statechart PingPongDataAnalyticsBehavior init Preprocess {
	/* The statechart specifies the behavior of this thing. Since this thing is responsible for DAML, its behavior can be modeled via a Finite-State Machine (statechart) that has four states: preprocess, train, ready and predict. Initially, the Preprocess state is necessary to do the data preparation. */
	
	on entry print "Ping Pong Data Analytics Started!\n"
	state Preprocess {
		on entry do
		print "Ping Pong Data Analytics: Data Preprocessing\n"
		da_preprocess da1 /* This action carries out the actual data preprocessing / preparation. */
		end
		transition -> Train /* This leads to the transition of the state machine (statechart) to the next state, Train. */
	}
\end{lstlisting}
\caption{Part of the model instance of the smart ping-pong example}
\label{lst:SmartPingPong1}
\end{figure} 

\begin{figure}[!htbp]
\setlength{\fboxsep}{0pt}%
\setlength{\fboxrule}{0pt}%
\centering
  \lstset{language=Java, showspaces=false,
    showstringspaces=false, tabsize=2, breaklines=true,
    xleftmargin=5.0ex,
	basicstyle=\color{blue}\ttfamily\footnotesize\bfseries,
	commentstyle=\color{black}\ttfamily\footnotesize,
	frame=tb,
	linewidth=1.0\textwidth
}
\begin{lstlisting}
  state Train {
   on entry do
    print "Ping Pong Data Analytics: Training\n"
	da_train da1 /* This action performs the training of the DAML model. */
   end
   transition -> Ready /* Once the training is done, the thing shall switch to the Ready (or idle) state to simply keep waiting for the incoming queries. */
  }
				
  state Ready {
   on entry do
    print "Ping Pong Data Analytics: Ready for Prediction\n"
   end
   transition -> Predict
   event m: da_service?query
   /* As soon as a message is received on the da_service port, the thing must switch to the Predict state. */
   action do
    /* Additionally, the following actions must be taken. */
    client_ip_address = m.client_ip  /* First, the value of the message parameter, called client_ip needs to be stored in the thing property (local variable) client_ip_address. */ 
    client_code = m.client_code /* Second, the value of the message parameter, called client_code must be stored in the thing property (local variable) client_code. */
   end
  }

  state Predict {
	on entry do
	print "Ping Pong Data Analytics: Predicting\n"
	da_predict da1(client_ip_address, client_code) /* This action asks the DAML model to make a prediction. */
	if(prediction==false)
	da_service!prediction_negative() /* If the prediction is false, send the prediction_negative message to pingServer, stating that pingClient is not likely to be an attacker. */
	else
	da_service!prediction_positive() /* Otherwise, send the prediction_positive message to pingServer, stating that pingClient is prone to be an attacker.	*/	
	end
	transition -> Ready /* In any case, switch back to the Ready (i.e., idle) state. */
	on exit da_save da1 /* This optional action results in appending the prediction to the dataset (CSV file). */
  }	
 }
}
\end{lstlisting}
\caption{Part of the model instance of the smart ping-pong example (continued)}
\label{lst:SmartPingPong2}
\end{figure} 

Finally, the user documentation available in our Github repository \cite{ML-Quadrat} provides further details for creating the desired smart IoT services using our modeling tool, as well as for getting involved in the development of the prototype as a contributor.

\section{Validation and Evaluation}\label{validation}
Section \ref{introduction} set out the underlying research hypotheses that must be assessed and validate in this work. They lead to the following Research Questions (RQ): \textbf{RQ1.} Can we enhance software models in domain-specific MDSE with the capability to automatically produce and train ML models, and deal with them, while maintaining the feasibility of full source code generation? \textbf{RQ2.} Will enhancing the software models and integrating them with the ML models contribute to the performance leap of software development in the IoT domain and lead to a higher level of satisfaction of the practitioners who use the proposed approach?

RQ1 that is concerned with the feasibility of the proposed approach is assessed using a case study with two use case scenarios in Section \ref{validation-case-study}. The research method here involves the implementation, simulation and testing of working examples \cite{Newman1994}. Further, RQ2 is assessed through an empirical user evaluation with four external volunteers in Section \ref{validation-empirical-eval}. Finally, we discuss the possible threats to validity in Section \ref{threats-to-validity}.

\subsection{Case Study}\label{validation-case-study}

The selected use case scenarios are from the domain of IoT/CPS, specifically smart energy systems in smart homes. The residential building, which is the data source, is located in the United Kingdom (UK). The data are publicly available through the REFIT datasets \cite{Murray2015, REFIT}. We use the data from House/Building $1$ from this dataset, which is a single-family dwelling with two inhabitants (a couple). Various sensors have recorded different conditions in their environment over a period of $21$ months starting from October 2013. The parameters of interest here are the individual loads (i.e., active power measured in Watts) of the following electrical appliances, as well as the aggregate load, i.e., the total power consumption of the entire house. The samples are recorded at a frequency of $0.125$ Hz, i.e., once every $8$ seconds. They include the following loads: (i) fridge, (ii) freezer-1, (iii) freezer-2, (iv) washing machine, (v) dishwasher, (vi) computer, (vii) television site, (viii) electric heater, and (ix) washer dryer. 

Some electricity providers, especially those who possess smart grids may offer certain discounts if the electrical appliances with higher consumption levels are avoided during the peak hours. Let us assume, there exists a database server that reads the values of the smart meters periodically and stores them for various smart home and ambient assisted living use cases. In this case study, we consider a smart grid that is also granted access to read this database. For them, it is only important whether a certain high energy consuming appliance, e.g., the washer dryer has been turned on during the peak hours or not. The exact power consumption does not really matter. However, due to various reasons, such as sensor malfunctions, power or network outages, or database failures, one might be faced with several missing values in the database. There exist different approaches to imputation of missing values in time series data. In this work, we deploy ML models as explained below, in order to predict the state (ON/OFF) of the washer dryer when the data are missing. Nevertheless, if the numerical value of the missing items must be estimated, e.g., in order to improve the quality of predictions of the ML models for other missing values in the future, then regression can be used (see scenario 3 below).

\begin{figure}[htbp]
	\centerline{\includegraphics[width=\textwidth]{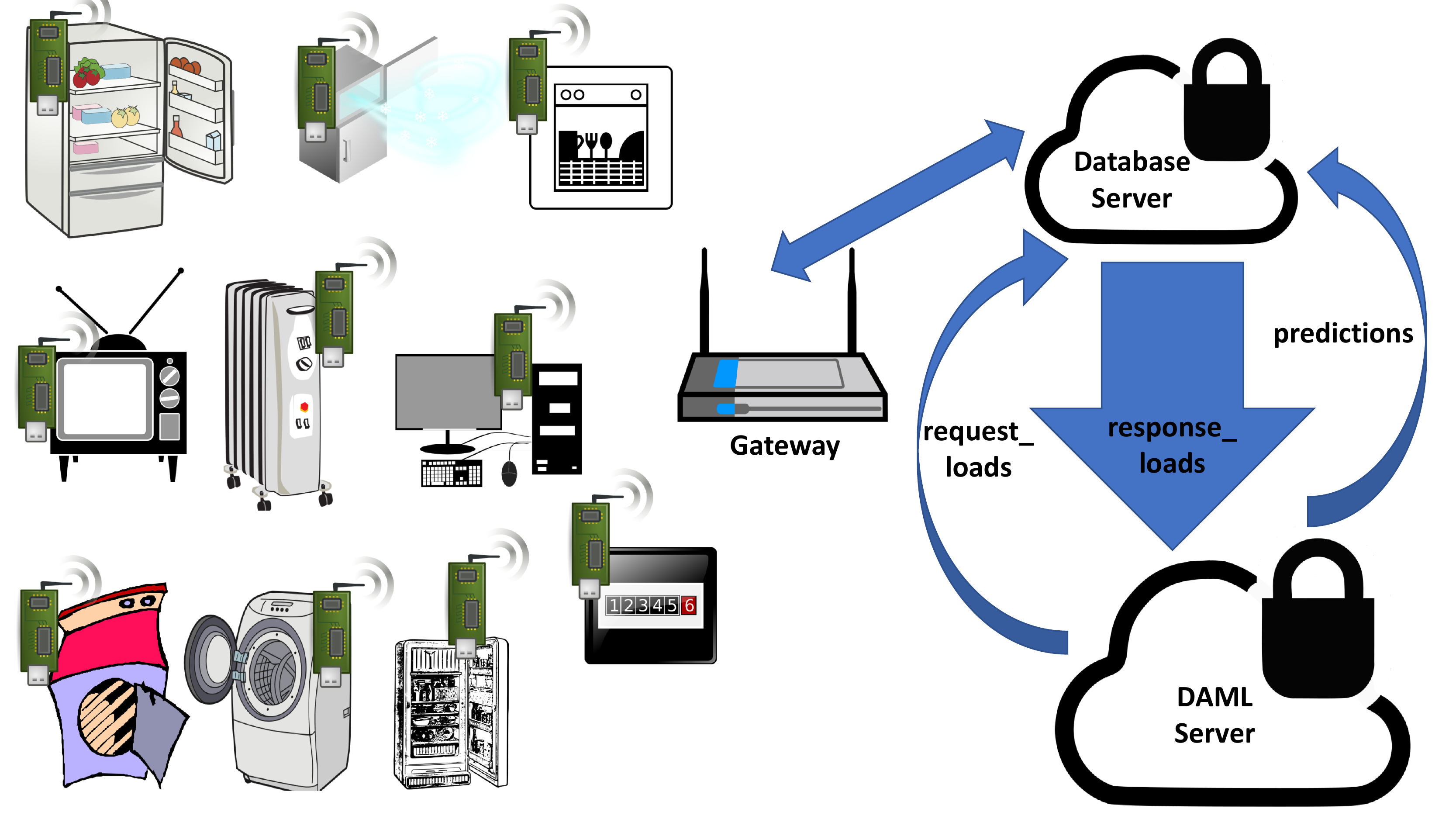}}
	\caption{The overall architecture of the target system for the case study}
	\label{fig:Usecase} 
\end{figure}

We consider four different scenarios (see below): (i) Classification, (ii) Clustering, (iii) Regression, and (iv) Black-box ML. In each case, the model instance comprises twelve \textit{things}: the nine electrical home appliances above, the said database server, as well as a meter that measures the aggregate load of the entire house, and a DAML server, which is responsible for the predictions of possible missing values in the database. In fact, in practice, the database server and the DAML server may or may not be deployed on the same physical node. Moreover, a gateway could be deployed at the entrance of the house. However, since the IoT advocates direct machine-to-machine communications and direct connections of the devices using their unique addresses \cite{Atzori+2010}, we skip the gateway in the present implementation. Figure \ref{fig:Usecase} illustrates the overall architecture of the system.

Each meter sends the active power of the corresponding appliance to the database server every eight seconds. Further, the DAML server sends a query to the database server in a periodic manner (e.g., once every $15$ minutes), asking for the latest sensor readings, i.e., the active powers of the nine appliances and the aggregate load of the house. Once the DAML server receives the response of the database server, which includes the ten requested values as message parameters, the DAML server can make a prediction about the missing values that are marked, for example, by \textit{NaN} in the database.

In the following, we illustrate the said scenarios. The full implementations of the respective model instances are included in the supplementary material of this work\footnote{See \url{https://doi.org/10.5281/zenodo.5501356}.}

\subsubsection*{Scenario 1: Classification (Supervised ML)}
We assume that the loads or active powers of the above-mentioned appliances are given together with the aggregate load of the house for time $t_i$. The task is to predict the binary status (ON/OFF) of the washer dryer at time $t_i$. The status of the washer dryer is used for the binary class labels of samples in the training dataset. We let the software model train the supervised ML model using $80\%$ of the available data. Thus, we keep $20\%$ of samples for testing the ML model. This is common practice in ML. For example, the Scikit-Learn \cite{Pedregosa+2011} library offers the \textit{train\_test\_split} method that is widely used \cite{Geron2019}. This method, by default, dedicates $25\%$ of the data to the test dataset unless another value is set for the \textit{test\_size} parameter. However, many practitioners simply follow the \textit{Pareto Principle} that is also called the $80/20$ rule, which states that in most cases, $80\%$ of effects come from $20\%$ of causes. Further research will be needed to see if a different split would yield better or worse results. Moreover, please note that we do not shuffle the data, i.e., we do not randomly split the data since they are sequential (namely time series) data and the order of the data instances matters. The supervised ML method deployed in this example is the Multi-Layer Perceptron (MLP) classifier from the Artificial Neural Networks (ANN) family with one hidden layer of size $100$, the \textit{Relu} activation function, the \textit{Adam} optimizer, the \textit{Sparse Categorical Cross Entropy} loss function, and the default values for the rest of the arguments/parameters of this ML method in the Scikit-Learn library.

The created software model instance has $545$ lines in the textual form. The model-to-code transformations generate $4,032$ Lines of Code (LoC) out of this. The generated source code contains $3,875$ lines of Java code and $157$ lines of Python code. The latter is responsible for the DAML functionalities and is seamlessly integrated with the Java code using the Java Process Builder API. Note that the scenarios below also exhibit the same number of LoC since we generate the APIs of the DAML library (in this case Scikit-Learn) and only the name of the ML method, as well as certain parameters/arguments change (but the number of the lines of code remain unchanged).

Furthermore, training the said ML model took $3,552$ seconds, and it performed with $100\%$ accuracy on the unseen test data (the ground truth comes from the mentioned open data, i.e., the REFIT datasets \cite{Murray2015, REFIT}). Typical ML performance metrics include but are not limited to accuracy, precision, recall and F1-Measure. In the case of binary classification, with the positive and negative classes, these are defined as follows:

\begin{equation}
  \label{eq:accuracy}
  Accuracy=\frac{TP+TN}{TP+TN+FP+FN}
\end{equation}
\begin{equation}
  \label{eq:precision}
  Precision=\frac{TP}{TP+FP}
\end{equation}

\begin{equation}
  \label{eq:recall}
  Recall=Sensitivity=\frac{TP}{TP+FN}
\end{equation}
\begin{equation}
  \label{eq:f1-measure}
  F1-Measure=\frac{2 . Precision . Recall}{Precision + Recall}
\end{equation}

In the equations above, TP, TN, FP, and FN are the True-Positive, True-Negative, False-Positive and False-Negative number of cases, respectively. 

In the said experiment, the other ML performance metrics, namely the precision, recall and F1-Measure were $99.9\%$, $100\%$ and $99.9\%$, respectively. The high performance was foreseeable given the fact that the ML task was not challenging for the MLP ANN classifier that is a highly capable one. In any case, the focus of this case study is not on measuring the performance of the ML methods since we only deploy the APIs of the target libraries for this purpose. The focus is rather on showing the feasibility of the proposed approach through the working examples. Hence, the reported performance figures in this section serve only for information purposes and are not supposed to contribute to the validation.

\subsubsection*{Scenario 2: Clustering (Unsupervised ML)}
Again, we assume that the loads or active powers of the above-mentioned appliances are given together with the aggregate load of the house for time $t_i$. Also, we have the same task, namely predicting whether the washer dryer is ON or OFF at time $t_i$. However, the training dataset this time has no labels for the data instances. This means, we do not know which sample in the training data belongs to the case when the washer dryer has been OFF and which one corresponds to the ON state of the washer dryer. The goal is to use the available data to train a clustering ML algorithm that can group the instances into two clusters: cluster A and cluster B. Cluster A, which we call it cluster 0 in the dataset corresponds to the OFF state of the washer dryer. In contrast, cluster B, which we call it cluster 1 in the dataset means the washer dryer has been ON. Note that 0 and 1 here are just the labels or names for the clusters and have no numerical interpretations. The unsupervised ML method deployed in this example is the K-Means clustering method with the values $2$ and $10$ provided for the arguments/parameters regarding the desired number of clusters and the random state of the algorithm, respectively. For the rest of the arguments/parameters of the method, the default values for this method in the Scikit-Learn library are considered. 

Furthermore, training the said clustering model took only $13$ seconds (extremely fast compared to the supervised model above), and it performed with $92\%$ accuracy on the unseen test data.

\begin{figure}[htbp]
	\centerline{\includegraphics[width=1.0\textwidth]{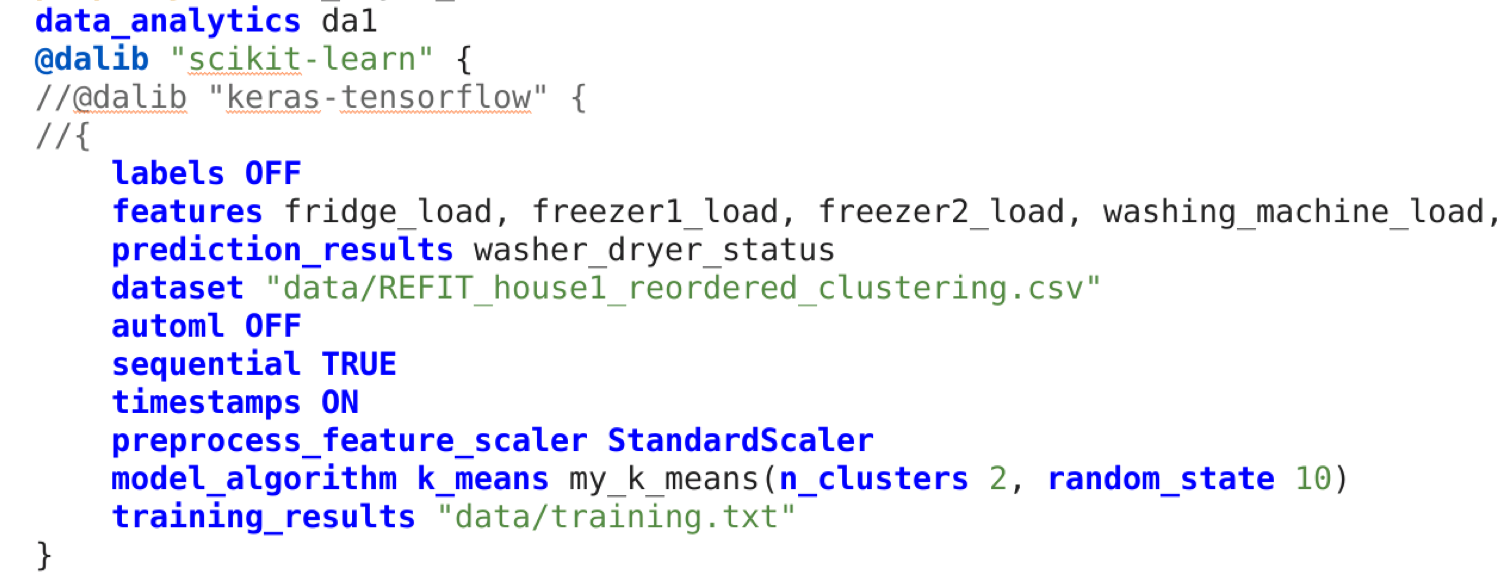}}
	\caption{The data analytics part of the software model instance in textual form}
	\label{fig:sample-model-full-mdse}
\end{figure}

\begin{figure}[htbp]
	\centerline{\includegraphics[width=1.0\textwidth]{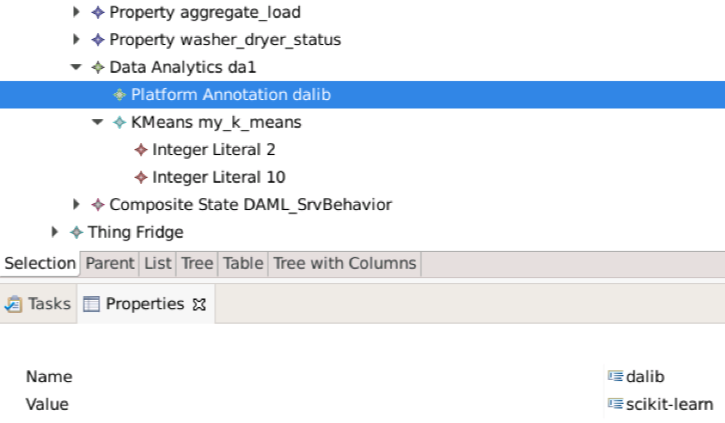}}
	\caption{The data analytics part of the software model instance in the EMF tree-based editor}
	\label{fig:sample-model-tree-based-editor}
\end{figure}

Figures \ref{fig:sample-model-full-mdse} and \ref{fig:sample-model-tree-based-editor} show a small part of the corresponding software model instance using the textual and the tree-based views of the concrete syntax in the Eclipse Modeling Framework (EMF).

\subsubsection*{Scenario 3: Regression (Supervised ML)}
This use case scenario is very similar to the first scenario above. However, instead of predicting the ON/OFF class labels, the task is to predict the numerical values of the active power of the washer dryer. We deployed the MLP ANN Regressor in Scikit-Learn.

For measuring the performance of regression, the typical error measures, Mean Absolute Error (MAE), also known as the L1-Norm, as well as the \textit{Mean Squared Error (MSE)}, also known as the \textit{L2-Norm} or the \textit{Euclidean Norm} are common choices. These are defined as follows:

\begin{equation}
  \label{eq:mae}
  MAE = \frac{1}{n}\displaystyle\sum_{i=1}^{n} \mid \hat{y}_i - y_i \mid
\end{equation}

\begin{equation}
  \label{eq:mse}
  MSE = \frac{1}{n}\displaystyle\sum_{i=1}^{n} ( \hat{y}_i - y_i)^2
\end{equation}

Here, $n$ is the number of data instances, $\hat{y}_i$ is the predicted numerical label for the $i-th$ data instance, and $y_i$ is the actual numerical label for this data instance.

The achieved MAE and MSE in the experiment above were $10.1$ and $29,962.1$, respectively.

\subsubsection*{Scenario 4: Black-box ML}
\begin{figure}[htbp]
	\centerline{\includegraphics[width=1.0\textwidth]{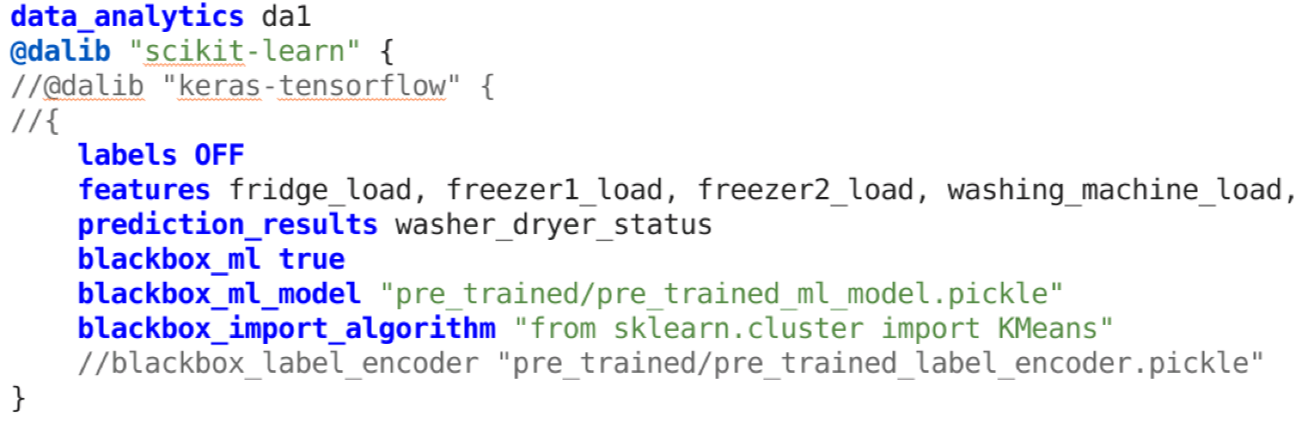}}
	\caption{The data analytics part of the software model instance that shows the black-box ML mode in textual form}
	\label{fig:sample-model-blackbox-ml}
\end{figure}

We train an unsupervised ML model without using the proposed approach. Thus, we develop the ML part manually. However, we use the same dataset. Then, we connect the pre-trained ML model to the software model using the black-box ML mode. The rest is the same as the unsupervised ML scenario above (including the performance). Figure \ref{fig:sample-model-blackbox-ml} demonstrates a small part of the respective software model instance.

\subsection{Empirical Evaluation}\label{validation-empirical-eval}
We ask four external experts in software engineering to use and evaluate our DSML through a number of experiments in a four-hour one-on-one video call over the Internet with short breaks in between. Two of them have a background in ML as well. Moreover, two of them work in academia and the other two work in the industry. Further, two out of four possess a PhD, whereas the rest have a Master's degree. Last but not least, they all belong to the age group of 25-39 years old, and one of them is a female. Table \ref{tab:evaluators-profiles} illustrates the self-reported levels of expertise of the evaluators in various fields, collected before carrying out the user experiments.

\begin{table}[htbp]
	\caption{The expertise levels of the evaluators}
	\begin{center}
		\begin{tabular}{|p{1cm}|p{3.5cm}|p{1cm}|p{1cm}|p{1.5cm}|}
			\hline	
			\textbf{Eval.\#} & \textbf{Software Engineering} & \textbf{DAML} & \textbf{MDSE} & \textbf{IoT/CPS} \\
			\hline
			\cellcolor{gray!25}1 & \cellcolor{gray!25}High & \cellcolor{gray!25}High & \cellcolor{gray!25}Low & \cellcolor{gray!25}Low \\
			\hline
		    2 & Medium & Low & Medium & Medium \\
			\hline
			\cellcolor{gray!25}3 & \cellcolor{gray!25}High & \cellcolor{gray!25}High & \cellcolor{gray!25}Low & \cellcolor{gray!25}Low \\
			\hline
			4 & Medium & Low & Medium & Medium \\
			\hline
		\end{tabular}
		\label{tab:evaluators-profiles}
	\end{center}
\end{table}

The evaluators are familiar with Java and Python programming. However, none of them has any background knowledge in the deployed DSMLs (neither in ThingML \cite{ThingML} nor in ours). During the four-hour sessions with the evaluators, we first deliver a 50-minute tutorial for using the proposed DSML, as well as the prior work on which we built our DSML, namely ThingML \cite{ThingML}. To this aim, we have already prepared a few samples, including a \lq{}HelloWorld\rq{} example. Moreover, we offer them our web-based prototype (see Section \ref{open-source-prototype}). We ask each evaluator to work on two tasks in three \textit{modes}: (a) Using pure manual software development (i.e., no MDSE); (b) Using the prior work, namely ThingML \cite{ThingML}; (c) Using the proposed DSML. We change the orders of the tasks, as well as the orders of the \textit{modes} for the four participants to avoid any bias and make the experiments fair. Both tasks are based on the case study set out in Section \ref{validation-case-study} above. However, in the first task, we ask the evaluator to use supervised ML (specifically classification as in the first scenario in Section \ref{validation-case-study}), whereas in the other task we ask for unsupervised ML (specifically clustering as in the third scenario in Section \ref{validation-case-study}). Recall that the use case scenario that was depicted in the case study in Section \ref{validation-case-study} involved $12$ \textit{things}. Implementing each of them gives the evaluator one point. An incomplete, but satisfactory implementation might result in $0.25$, $0.5$ or $0.75$ points, depending on the completeness and correctness of the implementation. Also, implementing the DAML component of each \textit{thing} (if it should have any) has one extra point (which may be granted only partially, depending on the status of the implementation as mentioned before). Table \ref{tab:evaluators-scores} summarizes the obtained points of the evaluators for all tasks and \textit{modes}. For each task, they have $75$ minutes time, which includes $25$ minutes per each \textit{mode}. During the experiments, they may maintain their access to their resources, e.g., the tutorials on the Internet and their own prior work, to make the experiments similar to the real-world practices of software developers and ML experts (e.g., data scientists). 

For the pure manual developments (i.e., in \textit{mode} a), we ask them to use Python for the ML part, with the APIs of the Scikit-Learn library and the ANN MLP classifier for the supervised task (i.e, task 1), as well as the K-Means clustering method for the unsupervised task (i.e., task 2). For the rest of their manual implementations, they are free to choose between Python and Java. However, in \textit{mode} b, they must deploy our web-based interface that offers the DSML and the code generators of ThingML (i.e., the prior work \cite{ThingML}) too, and implement the ML part manually in Python, so that their Python code can call the Java APIs of the generated Java code. Finally, in \textit{mode} c, no manual development will occur. They only use our web-based interface that offers our DSML and code generators to create their model instances. The full source code can be generated automatically. For the ML part, we currently generate Python code that is automatically integrated with the rest of the generated code.

\begin{table}[htbp]
	\caption{The scores of the 4 evaluators (Eval. \#1-4)}
	\begin{center}
		\begin{tabular}{|p{1.75cm}|p{1.5cm}|p{1.5cm}|p{1.5cm}|p{1.5cm}|p{0.75cm}|p{0.5cm}|}
			\hline	
			\textbf{Task-Mode} & \textbf{Eval. \#1} & \textbf{Eval. \#2} & \textbf{Eval. \#3} & \textbf{Eval. \#4} & \textbf{Total Score} & \textbf{Max.} \\
			\hline
			1-a & $1$ & $1$ & $2$ & $2$ & $6$ & $52$ \\
			\hline
            1-b & $0.5$ & $5.5$ & $3.25$ & $10.25$ & $19.5$ & $52$ \\
			\hline
			\cellcolor{gray!25}1-c & \cellcolor{gray!25}$2.25$ & \cellcolor{gray!25}$2$ & \cellcolor{gray!25}$5$ & \cellcolor{gray!25}$12$ & \cellcolor{gray!25}$21.25$ & \cellcolor{gray!25}$52$ \\
			\hline
			2-a & $0$ & $1$ & $2.25$ & $1$ & $4.25$ & $52$ \\
			\hline
			2-b & $2.25$ & $1.25$ & $4$ & $2$ & $9.5$ & $52$ \\
			\hline
			\cellcolor{gray!25}2-c & \cellcolor{gray!25}$2.25$ & \cellcolor{gray!25}$1.25$ & \cellcolor{gray!25}$5$ & \cellcolor{gray!25}$5$ & \cellcolor{gray!25}$13.5$ & \cellcolor{gray!25}$52$ \\
			\hline
		\end{tabular}
		\label{tab:evaluators-scores}
	\end{center}
\end{table}

As illustrated in Table \ref{tab:evaluators-scores}, using the proposed approach (see the rows 1-c and 2-c) has increased the scores of the evaluators, both compared to the prior work (see the rows 1-b and 2-b) and to the pure manual software development (see the rows 1-a and 2-a). The last column illustrates the total sum of the maximum possible scores for all of the evaluators, whereas the one before last column shows the total sum of the scores achieved by the evaluators in the experiments. Thus, we argue that the proposed approach may contribute to the improvement of the software development process efficiency. According to the experiments, the performance leap has been around $25\%$ on average, compared to the prior work (i.e., ThingML \cite{ThingML}) and around $236\%$ compared to the pure manual software development (see Section \ref{threats-to-validity}). We believe that the selected ML task was rather easy and only for one platform. One should be able to perceive a greater value in our proposed approach once heterogeneous IoT cloud and edge platforms need to be deployed. In the conducted experiments, many evaluators just started working on the DAML part from the very beginning. This should have resulted in a smaller difference between the productivity of software development in \textit{modes} b and c. Nevertheless, even $25\%$ productivity leap can still justify deploying the proposed approach.

Finally, we ask the opinions of the evaluators about their overall experience and satisfaction through a brief questionnaire at the end of the session. Compared to the prior work (ThingML \cite{ThingML}), two evaluators (\#1 and \#4) rated their level of satisfaction about the proposed approach as \textit{high}. Moreover, the other two evaluators chose the option \textit{medium}. The options were \textit{high}, \textit{medium} and \textit{low}. In contrast, when compared to pure manual software development, one of the evaluators selected the option \textit{low}. However, they emphasized that this answer is given the current exercises since the selected IoT platforms were not heterogeneous and it was rather easy for them to implement it manually. The other evaluators chose the answer options \textit{high}, \textit{medium} and again \textit{high} concerning this question. Hence, all in all, we argue that the proposed approach may contribute to the user experience and satisfaction of the practitioners.

\subsection{Discussion and Threats to Validity}\label{threats-to-validity}
The conducted experiments in Sections \ref{validation-case-study} and \ref{validation-empirical-eval} validated the first and the second hypotheses, respectively. First, we provided the proof-of-concept and showed the feasibility of enhancing MDSE models in the DSM methodology for developing IoT services with ML models if Artificial Intelligence (AI), in particular ML capabilities are required. Second, we verified empirically that the ML-enhanced software models used with the proposed approach can lead to performance leap for software development in the IoT domain and a higher satisfaction level of the practitioners compared to the prior model-driven work, namely ThingML \cite{ThingML} and the pure manual software development.

Recall that we claimed that DAML models (i.e., $DM$ in Equation \ref{eq:dm}) may affect the behavioral models of software systems (i.e., $B$ in Equation \ref{eq:sm}. This was shown formally through Equation \ref{eq:ai-enahnced-sm-behavior}. The way that the ML models, e.g., the ones corresponding to the above-mentioned use case scenarios for the case study in Section \ref{validation-case-study} can affect the behavioral models of software is through the use of the action type \textit{DA\_Predict} in the actions of the state machines (statecharts) that specify the behavioral models of the respective \textit{things}. The supplementary material of this paper\footnote{See \url{https://doi.org/10.5281/zenodo.5501356}.} shows the use of this action type for all of the depicted use case scenarios in Section \ref{validation-case-study}.

One key strength of this work for the SE community is expected to be that they gain access to the DAML methods and techniques out-of-the-box and can deploy them in their software models for the IoT. However, the major limitation is that ML methods cannot perform well if their hyperparameters are not tuned properly and/or the data that are used for training them are not prepared well. Therefore, more advanced AutoML features, e.g., concerning automated or semi-automated hyperparameter choices and tuning, as well as data preparation, e.g., for high-dimensional and non-i.i.d (independent and identically distributed) data are necessary (see Section \ref{conclusion-futurework}).

Further, a major advantage of this work for the DAML community is assumed to be that they can become involved in large-scale IoT projects easier as they will be able to work with the abstract software models that are easier to understand, adapt and use for them. Moreover, they may introduce any desired pre-trained ML model with any arbitrary architecture, learning algorithm and technique. This shall bring them a lot of flexibility as they will not be limited to the pre-defined options. However, the implication for them (as well as for the SE community) is that they have to be familiar with the DSML of the modeling tool and be willing to model their desired software using this DSML.

There exist a number of possible threats to the validity of the research results. First, we validated the first research hypothesis through a case study in Section \ref{validation-case-study}. We showed the feasibility of the proposed approach via a number of working examples with different use case scenarios. Although this is a well-established research method in engineering research (see, e.g., \cite{Newman1994}), we only had one overall case study domain (namely, smart energy data for smart home) and the selected case study and vertical application domain might not be representative enough for the entire domain of IoT/CPS. Thus, the generalized conclusions made here about the entire target domain might not be rigorous. 

Second, the empirical evaluation conducted in Section \ref{validation-empirical-eval} involved only four professionals. Consequently, the conclusions drawn may not hold for a larger sample group. In particular, the ideal research design should have involved randomized controlled experiments. However, our study was neither randomized nor had any control group. In contrast, we used convenience sampling and invited four independent, external volunteers to participate in our empirical evaluation. Further, the tasks chosen for the experiments were only two rather similar programming tasks with simple DAML requirements and no combination of heterogeneous resource-constrained IoT devices. This was due to the time and resource constraints for the experiments with the experts, but might be biased. Ideally, the tasks should have been more diverse and possibly more tasks would have been required, in order to be fair to different participants with different strengths. Additionally, we swapped the task and mode orders. However, we cannot rule out possible biases as a result of working on one task in a certain mode, e.g, using our DSML, and then in the following slot on the same task, but in a different mode, e.g, via pure manual software development. Also, it is clear that the time constraint may have an impact on the performance of evaluators in these tasks. For example, the manual task (namely, the \textit{a mode}) is expected to require more time than the tasks in the \textit{b mode} and the \textit{c mode}. Therefore, allotment of the same amount of time may not work ideally in all the modes. Finally, this was an exploratory user study/pilot study and a more rigorous evaluation with more evaluators is required in the future. Hence, the achieved preliminarily results might not be sufficient to perform a quantitative analysis.

\section{Conclusion and Future Work}\label{conclusion-futurework}
In this manuscript, we proposed a novel approach to marry the models in Artificial Intelligence (AI), specifically Machine Learning (ML), with the models in Software and Systems Engineering (SSE), particularly in Model-Driven Software Engineering (MDSE) following the Domain-Specific Modeling (DSM) methodology with full code generation. We showed how MDSE models can be integrated with ML models, thus become capable of producing and/or dealing with ML models. We concentrated on the Internet of Things (IoT) and Cyber-Physical Systems (CPS) domains, where both ML and MDSE are widely used. However, the proposed Domain-Specific Modeling Language (DSML), which is built based on the prior work in the literature, ThingML \cite{Morin+2017, Harrand+2016, Fleurey+2011, ThingML}, is not tied to any specific vertical application domain. Similar to the ThingML project \cite{ThingML}, we also supported full code generation in an automated manner through our ready-to-use model-to-code transformations. In addition to inheriting the code generators of ThingML \cite{ThingML}, we introduced a Python and Java code generator that can generate the APIs of the Scikit-Learn \cite{Pedregosa+2011} and Keras \cite{Chollet+2015} libraries and frameworks for ML. 

The two research hypothesis concerning the feasibility and the impact of the proposed approach were validated through the case study and the empirical user evaluation in Section \ref{validation}, respectively. It transpired that the proposed approach can lead to a higher performance and a better experience of the practitioner (e.g., software developer) for developing smart, data-driven IoT services. However, as stated in Section \ref{threats-to-validity}, a large-scale user study in the form of a randomized controlled experiment is required in the future.

The proposed approach has a number of limitations that can be addressed in future work. First, we supported supervised and unsupervised ML, whereas semi-supervised ML in which the data are only partially labeled is also desirable and beneficial in many use cases. Second, the pre-defined ML methods can be extended, e.g, with kernel methods, such as Support Vector Machines (SVM), Probabilistic Graphical Models (PGM), as well as more advanced ANN architectures, such as Long Short-Term Memory (LSTM) for Sequence-to-Sequence and End-to-End ML models. Third, more target platforms, programming languages and libraries can be supported. For instance, a pure Java code generator that uses the Java libraries WEKA or MOA (Massive Online Analysis) for DAML can be beneficial. Similarly, a pure Python code generator that does not have to mix the Python and Java codes for the IoT service might be advantageous for certain use cases, where Java might not be desired or useful. Last but not least, more advanced AutoML functionalities, e.g., concerning data preparation, as well as automated or semi-automated hyperparameter tuning will be very useful, in particular for software developers who might be novice in the field of DAML. 

Further, we implemented one specific variant of the proposed approach in Section \ref{proposed-approach}, where the DAML model may have an impact on the behavioral model of the software. However, it would be interesting to explore and realize other setups, e.g., where the DAML model might affect the structure of the software model, or even both the behavior and the structure. For instance, Pigem \cite{Pigem2013} studied how ML can be employed to \textit{learn} finite-state machines. Hence, there might be some potential in adopting such approaches and integrating them with the proposed approach to make the MDSE models even more intelligent. In fact, this would mean letting them learn the behavioral model of the software, in part or completely, on their own, using the existing data, instead of having the practitioner (i.e., the user of the DSML) specify it.

Finally, by enabling every \textit{thing} to possess one or more DAML components, we have enabled the modeling infrastructure for deploying edge analytics and federated learning. This paves the way for future work to provide a complete solution to supporting federated ML in the proposed DSML.

\section*{Data Availability}
The authors are committed to the open science initiative. Therefore, the entire research data are available as open data under the terms of the Creative Commons Attribution 4.0 International license at \url{https://doi.org/10.5281/zenodo.5501356}.

\begin{acknowledgements}
This work is partially funded by the German Federal Ministry for Education and Research (BMBF) through the Software Campus initiative (project ML-Quadrat). We are sincerely grateful to our external evaluators, Fatma Bozyigit from Izmir Bakircay University, Turkey, Burak Karaduman from University of Antwerp, Belgium, Andrei Mituca from DriotData UG, Germany, as well as the anonymous evaluator. The authors would like to also thank Stephan R\"ossler from Software AG and Marouane Sayih (alumnus of the Technical University of Munich) for their collaboration and support.
\end{acknowledgements}

\bibliographystyle{spbasic}      
\bibliography{refs}   

\begin{thebibliography}{51}
\providecommand{\natexlab}[1]{#1}
\providecommand{\url}[1]{{#1}}
\providecommand{\urlprefix}{URL }
\expandafter\ifx\csname urlstyle\endcsname\relax
  \providecommand{\doi}[1]{DOI~\discretionary{}{}{}#1}\else
  \providecommand{\doi}{DOI~\discretionary{}{}{}\begingroup
  \urlstyle{rm}\Url}\fi
\providecommand{\eprint}[2][]{\url{#2}}

\bibitem[{Bro(2008)}]{Broy2008}
 (2008) Time and causality in interactive distributed systems (lecture slides).
  \url{https://www5.in.tum.de/~huckle/Broy.pdf}, accessed on 2021-09-06

\bibitem[{ISO(2011)}]{ISO-IEC-IEEE-42010-2011}
 (2011) {ISO/IEC/IEEE 42010:2011 Systems and software engineering —
  Architecture description}. Standard, ISO / IEC / IEEE,
  \urlprefix\url{https://www.iso.org/standard/50508.html}

\bibitem[{Abadi et~al(2015)Abadi, Agarwal, Barham, Brevdo, Chen, Citro,
  Corrado, Davis, Dean, Devin, Ghemawat, Goodfellow, Harp, Irving, Isard, Jia,
  Jozefowicz, Kaiser, Kudlur, Levenberg, Man\'{e}, Monga, Moore, Murray, Olah,
  Schuster, Shlens, Steiner, Sutskever, Talwar, Tucker, Vanhoucke, Vasudevan,
  Vi\'{e}gas, Vinyals, Warden, Wattenberg, Wicke, Yu, and Zheng}]{Abadi+2015}
Abadi M, Agarwal A, Barham P, Brevdo E, Chen Z, Citro C, Corrado GS, Davis A,
  Dean J, Devin M, Ghemawat S, Goodfellow I, Harp A, Irving G, Isard M, Jia Y,
  Jozefowicz R, Kaiser L, Kudlur M, Levenberg J, Man\'{e} D, Monga R, Moore S,
  Murray D, Olah C, Schuster M, Shlens J, Steiner B, Sutskever I, Talwar K,
  Tucker P, Vanhoucke V, Vasudevan V, Vi\'{e}gas F, Vinyals O, Warden P,
  Wattenberg M, Wicke M, Yu Y, Zheng X (2015) {TensorFlow}: Large-scale machine
  learning on heterogeneous systems. \urlprefix\url{http://tensorflow.org/},
  software available from tensorflow.org

\bibitem[{Atzori et~al(2010)Atzori, Iera, and Morabito}]{Atzori+2010}
Atzori L, Iera A, Morabito G (2010) The internet of things: A survey. Computer
  Networks 54(15):2787 -- 2805

\bibitem[{de~Balle~Pigem(2013)}]{Pigem2013}
de~Balle~Pigem B (2013) Learning finite-state machines -- statistical and
  algorithmic aspects. PhD thesis, Universitat Polit`ecnica de Catalunya,
  Spain, \url{https://borjaballe.github.io/other/phdthesis.pdf}

\bibitem[{Berners-Lee and Hendler(2001)}]{Berners-LeeHendler2001}
Berners-Lee T, Hendler J (2001) Publishing on the semantic web. Nature
  410:1023--4, \doi{10.1038/35074206}

\bibitem[{Berthold et~al(2009)Berthold, Cebron, Dill, Gabriel, K\"{o}tter,
  Meinl, Ohl, Thiel, and Wiswedel}]{Berthold+2009}
Berthold MR, Cebron N, Dill F, Gabriel TR, K\"{o}tter T, Meinl T, Ohl P, Thiel
  K, Wiswedel B (2009) {KNIME - the Konstanz Information Miner: Version 2.0 and
  Beyond}. SIGKDD Explor Newsl 11(1):26--31, \doi{10.1145/1656274.1656280},
  \urlprefix\url{http://doi.acm.org/10.1145/1656274.1656280}

\bibitem[{Bishop(2006)}]{Bishop2006}
Bishop CM (2006) Pattern Recognition and Machine Learning (Information Science
  and Statistics). Springer-Verlag, Berlin, Heidelberg

\bibitem[{Bishop(2013)}]{Bishop2013}
Bishop CM (2013) Model-based machine learning. Philosophical Transactions of
  the Royal Society A 371(1984):1–17,
  \doi{https://doi.org/10.1098/rsta.2012.0222}

\bibitem[{CERN(n/a)}]{CERN}
CERN (n/a) The birth of the web.
  \url{https://home.cern/science/computing/birth-web}, accessed on 2021-09-06

\bibitem[{Chen et~al(2015)Chen, Li, Li, Lin, Wang, Wang, Xiao, Xu, Zhang, and
  Zhang}]{Chen+2015}
Chen T, Li M, Li Y, Lin M, Wang N, Wang M, Xiao T, Xu B, Zhang C, Zhang Z
  (2015) Mxnet: A flexible and efficient machine learning library for
  heterogeneous distributed systems. \eprint{1512.01274}

\bibitem[{Chollet et~al(2015)}]{Chollet+2015}
Chollet F, et~al (2015) Keras. \url{https://keras.io}

\bibitem[{Combemale et~al(2020)Combemale, Kienzle, Mussbacher, Ali, Amyot,
  Bagherzadeh, Batot, Bencomo, Benni, Bruel, Cabot, Cheng, Collet, Engels,
  Heinrich, Jézéquel, Koziolek, Mosser, Reussner, and Wimmer}]{Benoit+2020}
Combemale B, Kienzle J, Mussbacher G, Ali H, Amyot D, Bagherzadeh M, Batot E,
  Bencomo N, Benni B, Bruel JM, Cabot J, Cheng B, Collet P, Engels G, Heinrich
  R, Jézéquel JM, Koziolek A, Mosser S, Reussner R, Wimmer M (2020) A
  hitchhiker’s guide to model-driven engineering for data-centric systems.
  IEEE Software PP, \doi{10.1109/MS.2020.2995125}

\bibitem[{DiNucci(1999)}]{DiNucci1999}
DiNucci D (1999) Fragmented future. Print 53(4):32

\bibitem[{DMG(2021)}]{DMG}
DMG (2021) {Data Mining Group (DMG)}. \url{http://dmg.org}, accessed:
  2021-03-09

\bibitem[{Fleurey et~al(2011)Fleurey, Morin, Solberg, and
  Barais}]{Fleurey+2011}
Fleurey F, Morin B, Solberg A, Barais O (2011) Mde to manage communications
  with and between resource-constrained systems. In: Whittle J, Clark T,
  K{\"u}hne T (eds) Model Driven Engineering Languages and Systems, Springer
  Berlin Heidelberg, Berlin, Heidelberg, pp 349--363

\bibitem[{Fouquet et~al(2012)Fouquet, Morin, Fleurey, Barais, Plouzeau, and
  Jezequel}]{Fouquet+2012}
Fouquet F, Morin B, Fleurey F, Barais O, Plouzeau N, Jezequel JM (2012) A
  dynamic component model for cyber physical systems. In: Proceedings of the
  15th ACM SIGSOFT Symposium on Component Based Software Engineering,
  Association for Computing Machinery, New York, NY, USA, CBSE '12, p
  135–144, \doi{10.1145/2304736.2304759},
  \urlprefix\url{https://doi.org/10.1145/2304736.2304759}

\bibitem[{Geisberger and Broy(2014)}]{GeisbergerBroy2014}
Geisberger E, Broy M (eds)  (2014) Living in a networked world. {I}ntegrated
  research agenda {Cyber-Physical Systems (agendaCPS)}. acatech STUDY, Herbert
  Utz Verlag, Munich, Germany

\bibitem[{Greer et~al(2019)Greer, Burns, Wollman, and Griffor}]{Greer+2019}
Greer C, Burns M, Wollman D, Griffor E (2019) Cyber-physical systems and
  internet of things. \doi{https://doi.org/10.6028/NIST.SP.1900-202}

\bibitem[{GreyCat(2018)}]{GreyCat}
GreyCat (2018) {Next-Gen Live Analytics using Temporal Graph}.
  \url{https://github.com/datathings/greycat}, accessed: 2021-09-02

\bibitem[{Géron(2019)}]{Geron2019}
Géron A (2019) Hands-On Machine Learning with Scikit-Learn, Keras, and
  TensorFlow. O'Reilly Media, CA 95472, USA

\bibitem[{Harrand et~al(2016)Harrand, Fleurey, Morin, and Husa}]{Harrand+2016}
Harrand N, Fleurey F, Morin B, Husa KE (2016) {ThingML: A} language and code
  generation framework for heterogeneous targets. In: Proceedings of the
  ACM/IEEE 19th International Conference on Model Driven Engineering Languages
  and Systems, MODELS '16

\bibitem[{Hartmann et~al(2017)Hartmann, Moawad, Fouquet, and
  Le~Traon}]{Hartmann+2017}
Hartmann T, Moawad A, Fouquet F, Le~Traon Y (2017) The next evolution of mde: A
  seamless integration of machine learning into domain modeling. In: 2017
  ACM/IEEE 20th International Conference on Model Driven Engineering Languages
  and Systems (MODELS), pp 180--180, \doi{10.1109/MODELS.2017.32}

\bibitem[{Hartmann et~al(2018)Hartmann, Fouquet, Moawad, Rouvoy, and
  Traon}]{Hartmann+2018}
Hartmann T, Fouquet F, Moawad A, Rouvoy R, Traon YL (2018) Greycat: Efficient
  what-if analytics for data in motion at scale. \eprint{1803.09627}

\bibitem[{Hartmann et~al(2019)Hartmann, Moawad, Fouquet, and
  Le~Traon}]{Hartmann+2019}
Hartmann T, Moawad A, Fouquet F, Le~Traon Y (2019) The next evolution of mde: a
  seamless integration of machine learning into domain modeling. Software and
  System Modeling (SoSyM) 18:1285–1304,
  \doi{https://doi.org/10.1007/s10270-017-0600-2}

\bibitem[{HEADS(2015)}]{HEADS}
HEADS (2015) {Heterogeneous and Distributed Services for the Future Computing
  Continuum}. \url{https://cordis.europa.eu/project/id/611337}, accessed:
  2021-09-01

\bibitem[{Helwig et~al(2015)Helwig, Pignanelli, and Schütze}]{Helwig+2015}
Helwig N, Pignanelli E, Schütze A (2015) Condition monitoring of a complex
  hydraulic system using multivariate statistics. In: 2015 IEEE International
  Instrumentation and Measurement Technology Conference (I2MTC) Proceedings, pp
  210--215, \doi{10.1109/I2MTC.2015.7151267}

\bibitem[{Kelly and Tolvanen(2008)}]{KellyTolvanen2008}
Kelly S, Tolvanen JP (2008) Domain-Specific Modeling: Enabling Full Code
  Generation, 1st edn. Wiley

\bibitem[{Leskovec et~al(2014)Leskovec, Rajaraman, and Ullman}]{Leskovec+2014}
Leskovec J, Rajaraman A, Ullman JD (2014) Mining of Massive Datasets, 2nd edn.
  Cambridge University Press, USA, \urlprefix\url{http://www.mmds.org}

\bibitem[{Minka et~al(2018)Minka, Winn, Guiver, Zaykov, Fabian, and
  Bronskill}]{InferNet}
Minka T, Winn JM, Guiver JP, Zaykov Y, Fabian D, Bronskill J (2018) {Infer.NET
  0.3}. Microsoft Research Cambridge, \url{http://dotnet.github.io/infer},
  accessed: 2020-09-08

\bibitem[{ML-Quadrat(2020)}]{ML-Quadrat}
ML-Quadrat (2020) {ML2}. \url{https://github.com/arminmoin/ML-Quadrat},
  accessed: 2020-09-12

\bibitem[{Moin et~al(2018)Moin, R{\"{o}}ssler, and G{\"{u}}nnemann}]{Moin+2018}
Moin A, R{\"{o}}ssler S, G{\"{u}}nnemann S (2018) Thingml+: Augmenting
  model-driven software engineering for the internet of things with machine
  learning. In: Hebig R, Berger T (eds) Proceedings of {MODELS} 2018 Workshops,
  co-located with {ACM/IEEE} 21st International Conference on Model Driven
  Engineering Languages and Systems {(MODELS} 2018), Copenhagen, Denmark,
  October, 14, 2018, CEUR-WS.org, {CEUR} Workshop Proceedings, vol 2245, pp
  521--523, \urlprefix\url{http://ceur-ws.org/Vol-2245/mde4iot\_paper\_5.pdf}

\bibitem[{Moin et~al(2020)Moin, R{\"{o}}ssler, Sayih, and
  G{\"{u}}nnemann}]{Moin+2020}
Moin A, R{\"{o}}ssler S, Sayih M, G{\"{u}}nnemann S (2020) From things'
  modeling language (thingml) to things' machine learning (thingml2). In:
  Guerra E, Iovino L (eds) {MODELS} '20: {ACM/IEEE} 23rd International
  Conference on Model Driven Engineering Languages and Systems, Virtual Event,
  Canada, 18-23 October, 2020, Companion Proceedings, {ACM}, pp 19:1--19:2,
  \doi{10.1145/3417990.3420057}

\bibitem[{Morin et~al(2016)Morin, Fleurey, Husa, and Barais}]{Morin+2016}
Morin B, Fleurey F, Husa KE, Barais O (2016) A generative middleware for
  heterogeneous and distributed services. In: 2016 19th International ACM
  SIGSOFT Symposium on Component-Based Software Engineering (CBSE), pp
  107--116, \doi{10.1109/CBSE.2016.12}

\bibitem[{Morin et~al(2017)Morin, Harrand, and Fleurey}]{Morin+2017}
Morin B, Harrand N, Fleurey F (2017) Model-based software engineering to tame
  the iot jungle. IEEE Software 34(1):30--36, \doi{10.1109/MS.2017.11}

\bibitem[{Murray(2015)}]{Murray2015}
Murray D (2015) A data management platform for personalised real-time energy
  feedback. Proc 8th Int Conf Energy Efficiency Domestic Appl Lighting (EEDAL)
  pp 1--15

\bibitem[{Newman(1994)}]{Newman1994}
Newman W (1994) A preliminary analysis of the products of {HCI} research, using
  pro forma abstracts. In: Proceedings of the SIGCHI Conference on Human
  Factors in Computing Systems, Association for Computing Machinery, New York,
  NY, USA, CHI '94, p 278–284

\bibitem[{ONNX(2021)}]{ONNX}
ONNX (2021) {Open Neural Network Exchange}. \url{https://github.com/onnx},
  accessed: 2021-03-09

\bibitem[{Papatheocharous et~al(2013)Papatheocharous, Axelsson, and
  Andersson}]{Papatheocharous+2013}
Papatheocharous E, Axelsson J, Andersson J (2013) Issues and challenges in
  ecosystems for federated embedded systems. In: Proceedings of the First
  International Workshop on Software Engineering for Systems-of-Systems, pp
  21--24, \doi{10.1145/2489850.2489854}

\bibitem[{Paszke et~al(2017)Paszke, Gross, Chintala, Chanan, Yang, DeVito, Lin,
  Desmaison, Antiga, and Lerer}]{Paszke+2017}
Paszke A, Gross S, Chintala S, Chanan G, Yang E, DeVito Z, Lin Z, Desmaison A,
  Antiga L, Lerer A (2017) Automatic differentiation in pytorch. In: NIPS-W

\bibitem[{Pedregosa et~al(2011)Pedregosa, Varoquaux, Gramfort, Michel, Thirion,
  Grisel, Blondel, Prettenhofer, Weiss, Dubourg, Vanderplas, Passos,
  Cournapeau, Brucher, Perrot, and Duchesnay}]{Pedregosa+2011}
Pedregosa F, Varoquaux G, Gramfort A, Michel V, Thirion B, Grisel O, Blondel M,
  Prettenhofer P, Weiss R, Dubourg V, Vanderplas J, Passos A, Cournapeau D,
  Brucher M, Perrot M, Duchesnay E (2011) Scikit-learn: Machine learning in
  {P}ython. Journal of Machine Learning Research 12:2825--2830

\bibitem[{PFA(2021)}]{PFA}
PFA (2021) {Portable Format for Analytics}.
  \url{http://dmg.org/pfa/index.html}, accessed: 2021-03-09

\bibitem[{Pivarski et~al(2016)Pivarski, Bennett, and Grossman}]{Pivarski+2016}
Pivarski J, Bennett C, Grossman RL (2016) Deploying analytics with the portable
  format for analytics (pfa). In: Proceedings of the 22nd ACM SIGKDD
  International Conference on Knowledge Discovery and Data Mining, Association
  for Computing Machinery, New York, NY, USA, KDD '16, p 579–588,
  \doi{10.1145/2939672.2939731}

\bibitem[{PMML(2021)}]{PMML}
PMML (2021) {Predictive Model Markup Language}.
  \url{http://dmg.org/pmml/v4-4-1/GeneralStructure.html}, accessed: 2021-03-09

\bibitem[{RapidMiner(n/a)}]{RapidMiner}
RapidMiner (n/a) {Depth for Data Scientists, Simplified for Everyone Else}.
  \url{https://rapidminer.com}, accessed: 2021-09-08

\bibitem[{REFIT(2015)}]{REFIT}
REFIT (2015) {REFIT datasets}. \url{https://www.refitsmarthomes.org/datasets/},
  accessed: 2020-09-01

\bibitem[{Schaetz(2014)}]{Schaetz2014}
Schaetz B (2014) The role of models in engineering of cyber-physical systems
  – challenges and possibilities. In: CPS20: CPS 20 years from now - visions
  and challenges, CPS Week

\bibitem[{TensorBoard(n/a)}]{TensorBoard}
TensorBoard (n/a) {TensorFlow's visualization toolkit}.
  \url{https://www.tensorflow.org/tensorboard}, accessed: 2021-09-08

\bibitem[{{Theano Development Team}(2016)}]{Theano2016}
{Theano Development Team} (2016) {Theano: A {Python} framework for fast
  computation of mathematical expressions}. arXiv e-prints abs/1605.02688,
  \urlprefix\url{http://arxiv.org/abs/1605.02688}

\bibitem[{Things Modeling Language(2015)}]{ThingML}
Things Modeling Language (2015) {ThingML}.
  \url{https://github.com/TelluIoT/ThingML}, accessed: 2020-04-29

\bibitem[{Wang and Yeung(2020)}]{WangYeung2020}
Wang H, Yeung DY (2020) A survey on bayesian deep learning. ACM Comput Surv
  53(5), \doi{10.1145/3409383}, \urlprefix\url{https://doi.org/10.1145/3409383}

\end{thebibliography}

%
%
\end{document}